\documentclass[twocolumn,useAMS,usenatbib]{mn2e}
\usepackage{graphicx,natbib}
\usepackage{url}
\usepackage{natbib}
\title{Estimating the Redshift Distribution of Faint Galaxy Samples}

\author[Lima et al.]{
Marcos Lima$^{1,2}$\thanks{{\tt mvlima@uchicago.edu}},
Carlos E. Cunha$^{2,3}$,
Hiroaki Oyaizu$^{2,3}$,
Joshua Frieman$^{2,3,4}$,
\newauthor
Huan Lin$^{4}$,
Erin S. Sheldon$^{5}$
\\
${}^{1}$Department of Physics, University of Chicago, Chicago, IL 60637 \\
${}^{2}$Kavli Institute for Cosmological Physics, University of Chicago, Chicago, IL 60637 \\
${}^{3}$Department of Astronomy and Astrophysics, University of Chicago, Chicago, IL 60637 \\
${}^{4}$Center for Particle Astrophysics, Fermi National Accelerator Laboratory, Batavia, IL 60510 \\
${}^{5}$Center for Cosmology and Particle Physics and Department of Physics, New York University, New York, NY 10003 \\
}

\date{\today}

\begin{document}
\maketitle

\begin{abstract}
We present an empirical method for estimating the underlying redshift 
distribution $N(z)$ of galaxy photometric samples from photometric 
observables. The method does not rely on photometric redshift (photo-z) 
estimates for individual galaxies, which typically suffer from biases. 
Instead, it assigns weights to galaxies in a spectroscopic subsample 
such that the weighted distributions of photometric observables 
(e.g., multi-band magnitudes) match the corresponding distributions 
for the photometric sample. The weights are estimated using a 
nearest-neighbor technique that ensures stability in sparsely populated 
regions of color-magnitude space. The derived weights are then summed
in redshift bins to create the redshift distribution. We apply this 
weighting technique to data from the Sloan Digital Sky Survey as well as 
to mock catalogs for the Dark Energy Survey, and compare the results to 
those from the estimation of photo-z's derived by a neural network 
algorithm. We find that the weighting method accurately recovers the 
underlying redshift distribution, typically better than the photo-z 
reconstruction, provided the spectroscopic subsample spans the range 
of photometric observables covered by the photometric sample. 
\end{abstract}

\vspace{-0.3in}

\section{Introduction} \label{sec:int}

On-going, wide-field surveys 
are delivering photometric galaxy samples of unprecedented scale. 
Optical and near-infrared surveys planned for the next 
decade will increase the 
sizes of such samples by an order of magnitude. Much of the utility 
of these samples for astronomical and cosmological studies rests on 
knowledge of the redshift distributions of the galaxies they contain. 
For example, surveys aimed at probing dark energy via clusters, 
weak lensing, and baryon acoustic oscillations (BAO) will rely on the 
ability to coarsely bin galaxies by redshift, enabling approximate 
distance-redshift measurements as well as study of the growth of 
density perturbations. The power of these surveys to constrain 
cosmological parameters will be limited in part by the accuracy 
with which the galaxy redshift distributions can be determined
\citep{HutKimKraBro04,HutTakBerJai06,ZhaKno06,Zha06,MaHuHut06,LimHu07}.

Photometric redshifts (photo-z's, denoted $z_{\rm phot}$ below) 
-- approximate estimates of  
galaxy redshifts based on their broad-band 
photometric observables, e.g., magnitudes or colors -- offer one 
technique for approaching this problem. 
Photo-z's have the advantage 
that they provide redshift estimates for each galaxy 
in a photometric catalog; such information is useful 
for certain studies \citep{Manetal07}. 
However, in many applications we 
do not need such galaxy-by-galaxy information -- instead, we only require 
an estimate of the redshift distribution of a sample of galaxies 
selected by some set of photometric observables. For example, 
cosmic shear weak lensing or angular BAO measurements rely 
on relatively coarse binning of galaxies in redshift, and it 
suffices to have an accurate estimate of the redshift distribution $N(z)$ 
for galaxies satisfying certain color or magnitude selection 
criteria \citep{she04,Sheetal07a,jain07}.  
Photo-z estimators are {\it not} typically designed to provide 
unbiased estimates of the redshift distribution:  
$N(z_{\rm phot})$ is biased by photo-z errors.

Although deconvolution \citep{pad05} or other techniques 
\citep{Sheth07} can be used to obtain 
improved estimates of the redshift distribution from 
photo-z measurements, this problem 
motivates 
the development of a method optimized to directly estimate the underlying 
redshift distribution $N(z)$ for a photometric sample. In addition 
to its direct utility, a precise, unbiased estimate 
of the redshift distribution is useful 
even for probes that do require individual galaxy redshifts, 
since it provides a template for characterizing photo-z errors.

In this paper we present an empirical technique to estimate $N(z)$ for
a photometric galaxy sample that is based upon matching 
the distributions of photometric observables  
of a spectroscopic subsample to those of the photometric sample.
The method assigns weights to galaxies in the spectroscopic 
subsample (hereafter denoted the training set, in analogy with 
machine-learning methods of photo-z estimation), so 
that the weighted distributions of observables for these galaxies 
match those of the photometric sample. The weight for 
each training-set galaxy is computed by comparing the local ``density''  
of training-set galaxies
in the multi-dimensional space of photometric 
observables to the density of the 
photometric sample in the same region. 
We estimate the densities using a nearest neighbor approach that 
ensures the density estimate is both local and stable in sparsely 
occupied regions of the space. 
The use of the nearest neighbors ensures optimal binning of the data, which 
minimizes the requisite size of the spectroscopic sample. 
After the training-set galaxy weights are derived, we sum them 
in redshift bins to estimate the redshift distribution. 

As we will show, this method provides a precise and nearly unbiased 
estimate of the underlying redshift distribution for a photometric sample 
and does not require photo-z estimates for individual galaxies. 
Moreover, the spectroscopic training set does {\it not} have to be 
representative of the photometric sample, 
in its distributions of magnitudes, colors, or redshift, 
for the method to work. 
We only require that the spectroscopic training set {\it cover}, even 
sparsely, the range of photometric observables spanned by the photometric 
sample. 
The method can be applied to different combinations of photometric 
observables that correlate with redshift -- in this paper, 
we confine our analysis to magnitudes and colors. 
In a companion paper (Cunha et al., in preparation), we compare this 
weighting technique to the deconvolution method of \cite{pad05} and show 
that the weights can be used to naturally regularize and improve the 
deconvolution.

The paper is organized as follows.
In \S~\ref{sec:cat} we present the simulated and real galaxy catalogs 
used to test the method. 
In \S~\ref{sec:wei} we describe the algorithm 
for the calculation of the weights of training-set galaxies using a nearest 
neighbor method. 
In \S~\ref{sec:qua} we define simple statistics to assess the quality
of the reconstructed distributions. 
In \S~\ref{sec:res} we present $N(z)$ 
estimates derived from the weighting method 
and compare with results using photo-z's
for individual galaxies derived from a neural-network algorithm. 
We discuss the results and present our conclusions and perspectives in 
\S~\ref{sec:dis}.
In Appendix~\ref{app:neu}, we provide a brief description of the neural
network photo-z algorithm that we use for comparison with the weighting 
method.

\section{Catalogs} \label{sec:cat}
We use two sets of catalogs to test the method. The first is based upon 
simulations of the Dark Energy Survey (DES). 
The second derives from photometry for galaxies in the 
Sloan Digital Sky Survey (SDSS). 
We describe them in turn.

\subsection{DES mock catalogs} \label{subsec:DESmock}

\begin{figure*}
  \begin{minipage}[t]{58mm}
    \begin{center}
      \resizebox{58mm}{!}{\includegraphics[angle=0]{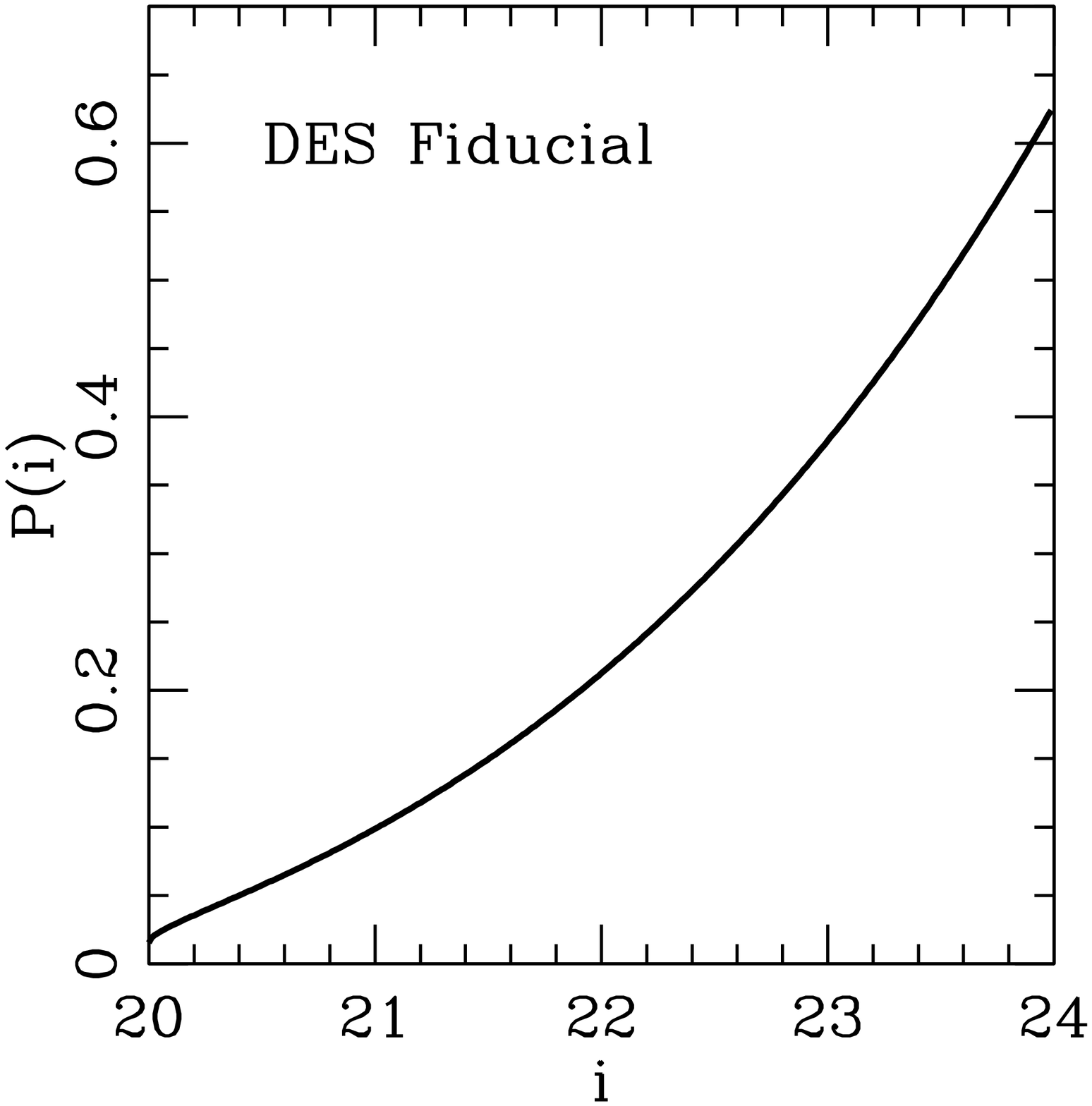}}
    \end{center}
  \end{minipage}
  \begin{minipage}[t]{58mm}
    \begin{center}
      \resizebox{58mm}{!}{\includegraphics[angle=0]{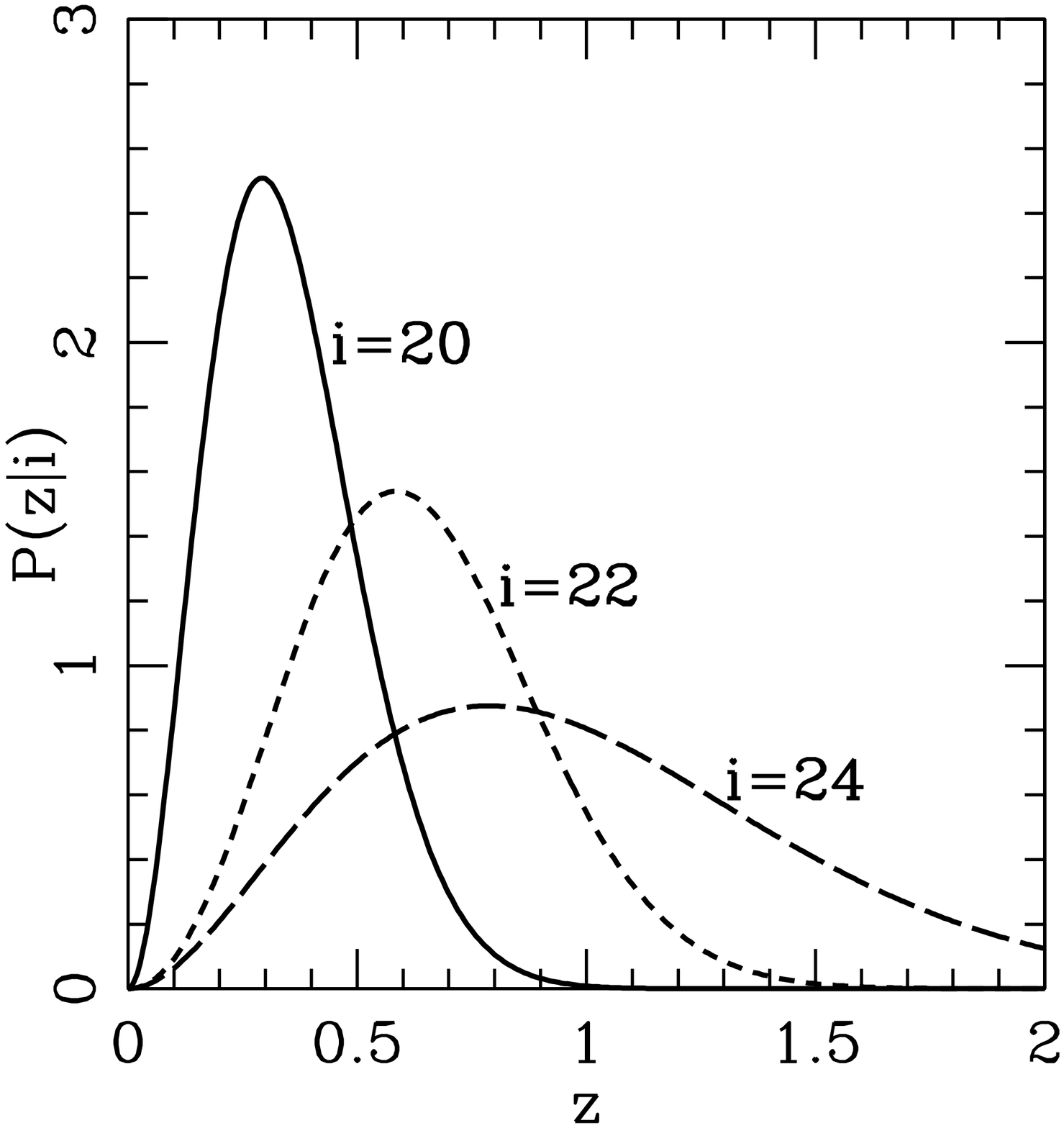}}
    \end{center}
  \end{minipage}
  \begin{minipage}[t]{58mm}
    \begin{center}
      \resizebox{58mm}{!}{\includegraphics[angle=0]{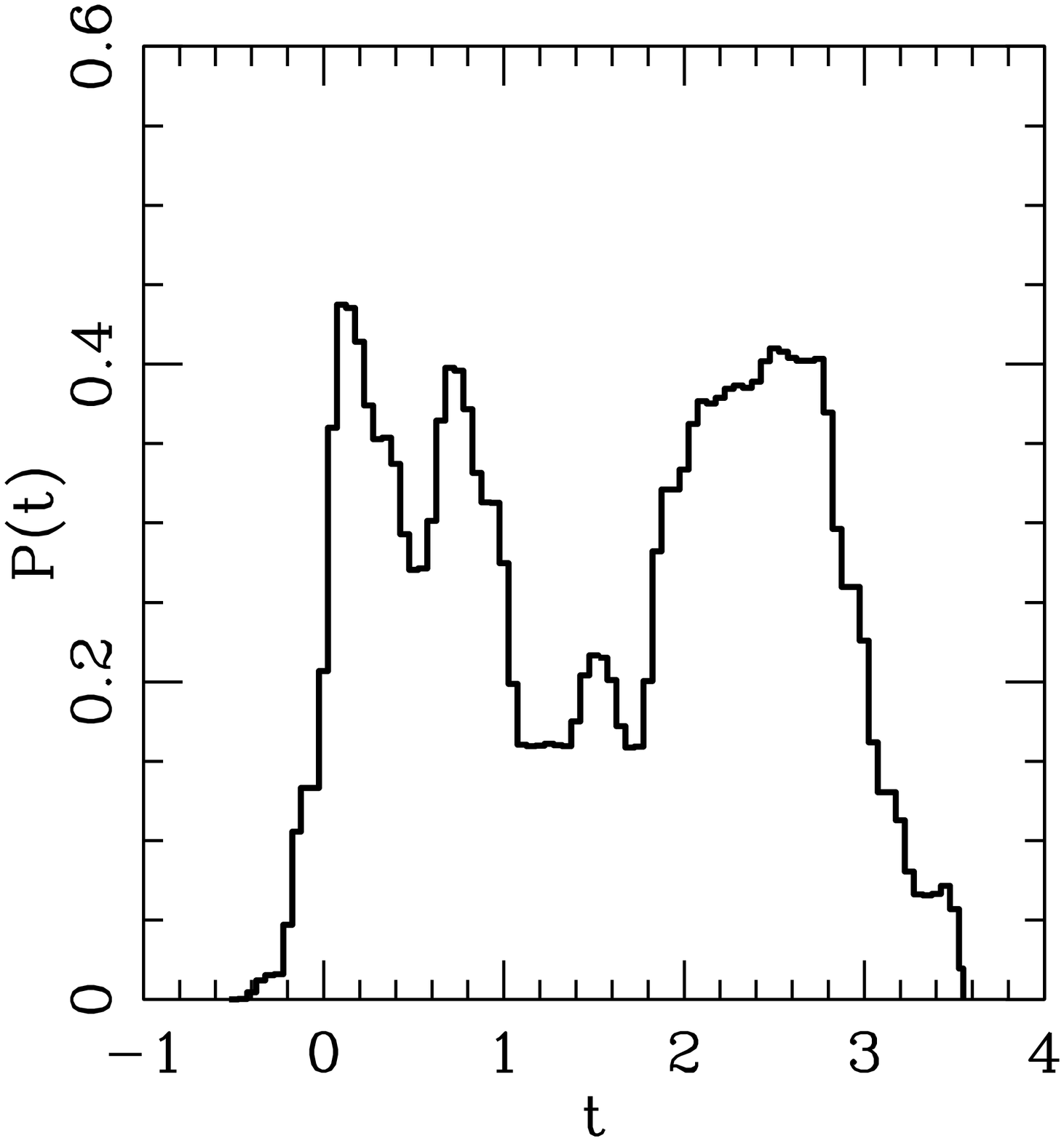}}
    \end{center}
  \end{minipage}
\caption{
Distributions of: $i$ magnitude ({\it left panel}); redshift $z$ given 
$i$-magnitude for $i=$ 20, 22, 24  
({\it middle panel}); and galaxy type $t$ ({\it right panel}) for the fiducial 
DES mock catalog. Lower (higher) values of $t$ correspond to early (late) 
spectral types, and the $t$ distribution shows evidence of bimodality.
}\label{fig:prob.DESmock.fid}
\end{figure*}

The Dark Energy Survey is a 5000 square degree survey in 5 optical passbands 
($grizY$) with an AB-magnitude limit of $i \approx 24$ (the 
approximate 10$\sigma$ limit for 
galaxies), to be carried out using 
a new camera on the CTIO Blanco 4-meter telescope. 
The goal of the survey is to measure the equation  of state of dark energy 
using several techniques: clusters of galaxies, weak lensing, BAO and 
supernovae. 
The DES optical survey will be complemented in the near-infrared by the VISTA 
Hemisphere Survey (VHS), an ESO Public Survey on the VISTA 4-meter telescope 
that 
will cover the survey area in three near-infra-red (NIR) bands ($JHKs$).
For simplicity we will only use the optical DES bands in our results
and analysis presented below.

Our fiducial simulated DES catalog contains 500,000 galaxies 
with redshift $z < 2$ and with $20 < i < 24$, and will serve
as the photometric set we will be attempting to recover.
The magnitude and redshift distributions were derived from the galaxy 
luminosity function measurements of \cite{lin99} and \cite{pol03}, 
while the galaxy Spectral Energy Distribution (SED) 
type distribution was obtained from measurements 
of the HDF-N/GOODS field \citep{cap04,wir04,cow04}. 
The galaxy colors were generated using the four \citet{col80} templates -- E, 
Sbc, Scd, Im -- extended to the UV and NIR using synthetic templates 
from \citet{bru93}. 
These templates are mapped to a galaxy SED type $t$ as 
$($E, Sbc, Scd, Im$)$ $\rightarrow t=(0,1,2,3)$.
To improve the sampling and coverage of color space, we created additional 
templates by interpolating between adjacent templates and by extrapolating 
from the E and Im templates, such that the SED type $t$ ranges
over $[-0.5,3.5]$ continuously, with $t=-0.5$ (3.5) corresponding
to very early-type (very late-type) galaxies.
The magnitude errors were modelled as sky-background dominated errors 
approximated as uncorrelated Gaussians.
This implementation of the DES mock catalog is similar to the one 
employed in \cite{Ban07}.

In order to vary the parameters of this fiducial DES catalog and to 
create spectroscopic training sets from it, we adopt an equivalent 
analytic description of this sample that is easier to work with.
The photometric sample can be fully specified by providing the 
distributions of $i$ magnitude, redshift
$z$, and SED type $t$. 
That is, the catalog can be constructed by repeated sampling from a 
probability distribution $P(i,z,t)$, since the ``pre-noise'' 
magnitudes in the other passbands are uniquely determined by these 
three quantities. 

We can write the probability $P(i,z,t)$ as a product of 
conditional probabilities
\begin{eqnarray}
P(i,z,t)=P(i)P(z|i)P(t|i,z) \, ,
\end{eqnarray}

\noindent where $P(i)$ is the probability that a galaxy in the sample has $i$-band magnitude $i$,
$P(z|i)$ is the probability that a galaxy of that magnitude has redshift $z$, 
and $P(t|i,z)$ is the probability that a galaxy of that magnitude and redshift 
has SED type $t$.  
The galaxy $i$-magnitudes, redshifts $z$ and SED types $t$ have the ranges 
specified above, i.e. the conditional probability distributions are 
truncated sharply at those values and normalized by
\begin{eqnarray}
\int_{20}^{24} P(i) di = \int_{0}^{2} P(z|i) dz = \int_{-0.5}^{3.5} P(t|i,z) dt =1 \, ,
\end{eqnarray}

\noindent which implies that $P(i,z,t)$ is properly normalized.
For the DES sample generated according to the observed luminosity 
function and SED type distributions 
noted above, we find that the magnitude and redshift distributions 
can be accurately parametrized by 
\begin{eqnarray}
P(i)&=& A \exp\left[ \left(\frac{i-20}{a}\right)^{0.5} \right] \, , 
\label{eqn:Pi} \\
P(z|i)&=& B z^2 \exp\left[-\left(\frac{z-z_d(i)}{\sigma_d(i)}\right)^2 \right] \, ,
\label{eqn:Pzi}
\end{eqnarray}

\noindent where the functions $z_d(i)$ and $\sigma_d(i)$ are defined by 
\begin{eqnarray}
z_d(i)&=& b_1+b_2(i-20)+b_3(i-20)^2 \, , \label{eqn:zd} \\
\sigma_d(i)&=& c_1+c_2(i-20)^{c_3}\, .
\label{eqn:sigd} 
\end{eqnarray}

\noindent Here $A=A(a)$ and $B=B(z_d,\sigma_d)$ are normalization factors 
determined once the constants $a$, $b_j$, $c_j$ are specified. 
For the photometric sample of the mock DES catalog, we find good fits with  
$a=0.29$, $(b_1,b_2,b_3)=(-0.2,0.75,-0.28)$, and 
$(c_1,c_2,c_3)=(0.39,0.012,3.2)$. 
We therefore use these parametric distributions 
to generate the mock DES samples for our analysis. 
The resulting analytic distributions 
are shown in the first two panels of Fig.~\ref{fig:prob.DESmock.fid}.

For simplicity, we assume that the SED type distribution is independent 
of magnitude and redshift,  
\begin{equation}
P(t|i,z)= P(t) \, ,
\label{eqn:Pt}
\end{equation}

\noindent and has the bimodal shape given in the third panel of 
Fig.~\ref{fig:prob.DESmock.fid},
which comes from the original construction of the catalog in terms of
luminosity functions and the HDF-N/GOODS type distribution.

In \S~\ref{sec:res} we explore how well
the redshift distribution $N(z)$ of the DES mock photometric sample created 
by this prescription, shown in the right panel of Fig. 
\ref{fig:all.z.dist.DESmock.tflat.igfdf} below,  
can be recovered from spectroscopic training
sets that have different $P(i,z,t)$ distributions from 
the photometric sample.

\subsection{SDSS Data Catalogs} \label{subsec:SDSScat}

\begin{figure*}
  \begin{minipage}[t]{180mm}
    \begin{center}
      \resizebox{180mm}{!}{\includegraphics[angle=0]{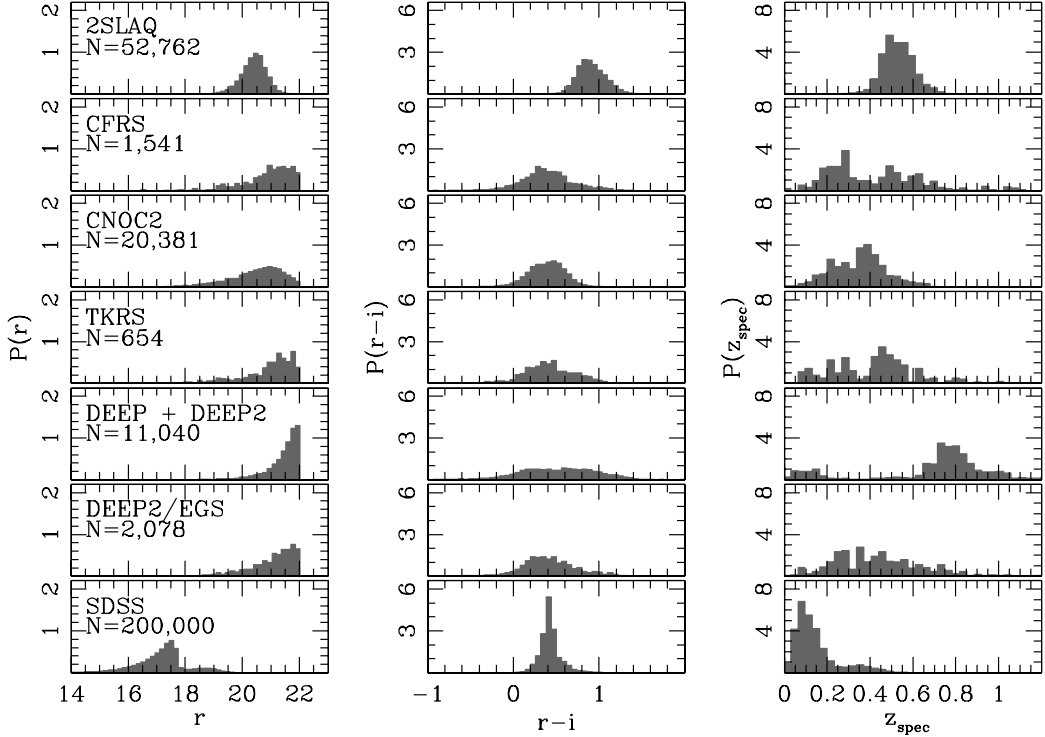}}
    \end{center}
  \end{minipage}
\caption{
Distributions of $r$ magnitude ({\it left panels}), 
$r-i$ color ({\it middle panels}), and 
spectroscopic redshift $z_{\rm spec}$ ({\it right panels}) for 
each spectroscopic catalog used with SDSS photometry. 
Also shown in the left panels are the total numbers of galaxies in each
spectroscopic sample, counting repeated objects.
}\label{fig:dndrz_real}
\end{figure*}

While the mock catalogs are useful for study of parameter dependencies 
and to gain insight into the efficiency and requirements of the $N(z)$ 
reconstruction method, they do not capture all the degeneracies and features 
of real catalogs. 

Therefore, we also test the weighting procedure using a combination of
 spectroscopic catalogs with SDSS DR6 photometry in $ugriz$ bands.  
The derived spectroscopic sample is similar to the 
one we used in constructing the DR6 galaxy photo-z catalog 
\citep{OyaLimCunLinFriShe08}, and we apply the same redshift 
quality and photometry cuts as were used there.
Together these catalogs contain 
$288,456$ galaxies with $r<22$. 
We use $200,000$ galaxies from the SDSS DR6 
main and LRG spectroscopic samples, 
$20,381$ from the Canadian Network for Observational Cosmology (CNOC) 
Field Galaxy Survey \citep[CNOC2;][]{yee00},  
$1,541$ from the Canada-France Redshift Survey \citep[CFRS;][]{lil95},  
$11,040$ color-selected galaxies 
from the Deep Extragalactic Evolutionary Probe \citep[DEEP;][]{deep2} and the 
DEEP2 surveys \citep{wei05}\footnote{{\tt http://deep.berkeley.edu/DR2/ }},
$2,078$ galaxies from a roughly flux-limited 
sample from the Extended Groth Strip in DEEP2 \citep[DEEP2/EGS;][]{dav07}, 
$654$ from the Team Keck Redshift Survey \citep[TKRS;][]{wir04}, and 
$52,762$ from the 
2dF-SDSS LRG and QSO Survey 
\citep[2SLAQ;][]{can06}
\footnote{{\tt http://lrg.physics.uq.edu.au/New\_dataset2/ }}. 
The numbers of galaxies used from each catalog are smaller than
those in \cite{OyaLimCunLinFriShe08}, because we cut the 
samples at $r<22$,  
as opposed to the $r<23$ limit adopted in that work. 
Also, the numbers above include repeat objects due to 
repeat imaging in the SDSS {\tt BestRuns} database, which was used 
to positionally match the galaxies.

In Fig.~\ref{fig:dndrz_real}, we show the distributions of $r$ magnitude,
$r-i$ color, and spectroscopic redshift
$z_{\rm spec}$ for each spectroscopic catalog. In combination, 
these data sets span a large range of magnitude, color, and redshift. 
We use these catalogs as a test case for 
our reconstruction methods of the redshift distribution below.
In Cunha et al. (in preparation), we use simulations to investigate the 
effectiveness of the weighted $N(z)$ estimation on the SDSS DR6 Photoz2 
sample described in \cite{OyaLimCunLinFriShe08}.

\section{The Weighting Method} \label{sec:wei}

The weighting method for reconstructing $N(z)$ for a photometric sample 
relies on the fact that a spectroscopic subsample of the galaxies with 
precisely measured redshifts is usually available. However, 
due to observational constraints, the spectroscopic subsample typically 
has different distributions of magnitudes, colors, and 
therefore redshift than the 
parent photometric sample, e.g., the spectroscopic sample may contain 
galaxies that are mostly much brighter than the flux limit of the 
photometric sample or the spectroscopic sample may be selected 
to lie within certain windows of color space.
The weighting technique compensates for this mismatch by weighting galaxies 
in the spectroscopic sample so that the weighted sample has the {\it same} 
distribution of photometric observables (colors, magnitudes) as the 
parent photometric sample. The key assumption behind the method is that 
two samples with identical distributions of photometric observables 
will have identical distributions of redshift $N(z)$, so that the 
redshift distribution for the weighted spectroscopic sample serves as 
an estimate of the redshift distribution for the photometric sample.
In \S~\ref{sec:dis}, we discuss the conditions that are required for this 
assumption to hold and the systematic errors that can arise for the 
$N(z)$ estimate if those 
conditions are not met.

In the remainder of this Section, we describe the construction of 
the weighting method. 

\subsection{Matching Distributions: Redshifts} 

We are interested in estimating $N(z)$ for a photometric set of interest. 
In practice, to estimate this distribution, we need to bin galaxies and
compute the binned redshift distribution. 
Consider first binning the photometric and spectroscopic samples by 
{\it non-overlapping} redshift bins, denoted $i$ below. 
We define the normalized redshift distributions in the $i^{\rm th}$ 
redshift bin $[z^i,z^i+\Delta z^i]$ 
in the photometric sample (superscript P) and in the spectroscopic 
training set (superscript T) as 
\begin{eqnarray}
P(z^i)^{\rm P} &\equiv& \frac{1}{N_{\rm tot}^{\rm P}}\frac{N(z^i)^{\rm P}}{\Delta z^i} = \frac{\rho(z^i)^{\rm P}}{N_{\rm tot}^{\rm P}}\,,\label{eqn:probP_def} \\
P(z^i)^{\rm T} &\equiv& \frac{1}{N_{\rm tot}^{\rm T}}\frac{N(z^i)^{\rm T}}{\Delta z^i} = \frac{\rho(z^i)^{\rm T}}{N_{\rm tot}^{\rm T}}\,,
\label{eqn:prob_z_def}
\end{eqnarray}

\noindent where $N_{\rm tot}$ is the total number of galaxies  
in each catalog, $N(z^i)$ is the number of galaxies in the $i^{\rm th}$ 
redshift bin, and we have defined
the redshift density
\begin{eqnarray}
\rho(z^i) \equiv \frac{N(z^i)}{\Delta z^i} \,.
\label{eqn:den_z_gen_def}
\end{eqnarray}
  
In general, $P(z^i)^{\rm T} \neq P(z^i)^{\rm P}$. Using 
weights, we would like to transform $P(z^i)^{\rm T}$ into 
a new distribution,  $P(z^i)^{\rm T}_{\rm wei}$, that 
provides an unbiased 
estimate of $P(z^i)^{\rm P}$, 
\begin{eqnarray}
P(z^i)^{\rm P}=\langle P(z^i)^T_{\rm wei} \rangle \, .
\label{eqn:probP_probTwei_equiv}
\end{eqnarray}

\noindent To accomplish this, we weight objects in the spectroscopic training 
set according to their local density in the space of photometric 
observables. 

To motivate the form of the weights, we first consider 
the idealized case of weighting directly in redshift. 
We seek a set of weights $W_{\alpha}$ for the  
spectroscopic training-set
galaxies indexed by $\alpha$ 
such that the redshift distribution of the weighted training set 
is given by
\begin{eqnarray}
P(z^i)^{\rm T}_{\rm wei} \Delta z^i \equiv
\frac{  \sum_{\alpha=1}^{N(z^i)^{\rm T}}{W_{\alpha}}      }
     {  \sum_{\alpha=1}^{N_{\rm tot}^{\rm T}}{W_{\alpha}} } \, ,
\label{eqn:p_wei_def}
\end{eqnarray}

\noindent where the sum in the numerator is over objects in the $i^{\rm th}$ 
redshift bin and that in 
the denominator is over all objects in the spectroscopic training set.
The weights can be normalized by 
\begin{eqnarray}
\sum_{\alpha=1}^{N_{\rm tot}^{\rm T}} W_{\alpha} = 1 \, .
\label{eqn:wei_norm}
\end{eqnarray}

Clearly, Eq.~(\ref{eqn:p_wei_def}) reduces to 
Eq.~(\ref{eqn:prob_z_def}) if all objects have the same weight.
We write the Eqs. above in terms of non-overlapping redshift bins
because that is what we will show in our results after computing
the galaxy weights $W_{\alpha}$.
Consider now an alternate estimate of $N(z)$ with  $overlapping$ bins
centered at individual galaxies, indexed by $\beta$, with corresponding
bin sizes $\Delta z_{\beta}$, possibly varying from galaxy to galaxy. 
All Eqs.~(\ref{eqn:probP_def}-\ref{eqn:p_wei_def}) apply after substituting 
indices $i$ by $\beta$.
For sufficiently narrow bins $\Delta z_{\beta}$, galaxies inside a given 
bin will be roughly indistinguishable and will have
approximately equal weights, labeled $W_{\beta}$. 
In this case, 
\begin{eqnarray}
\sum_{\alpha=1}^{N(z_{\beta})^{\rm T}}{W_{\alpha}} \sim N(z_{\beta})^{\rm T} W_{\beta}  \,.
\label{eqn:bin_approx}
\end{eqnarray}

\noindent Combining the results of 
Eqs.~(\ref{eqn:probP_def}), 
(\ref{eqn:probP_probTwei_equiv})-(\ref{eqn:bin_approx}) 
yields
\begin{eqnarray}
\frac{1}{N_{\rm tot}^{\rm P}}\frac{N(z_{\beta})^{\rm P}}{\Delta z_{\beta}}=
\frac{ N(z_{\beta})^{\rm T} W_{\beta}}{\Delta z_{\beta} } \,,
\end{eqnarray}

\noindent from which it follows that the idealized weights are 
given by
\begin{eqnarray}
W_{\beta} = \frac{1}{N_{\rm tot}^{\rm P}}\frac{\rho(z_{\beta})^{\rm P}}{\rho(z_{\beta})^{\rm T}} \,.
\label{eqn:wei_z_gen_def}
\end{eqnarray}

Of course, Eq.~(\ref{eqn:wei_z_gen_def}) is not useful in practice, 
since we do not know how to estimate the redshift density $\rho(z)^P$ 
for the photometric sample. 
In principle, one could do the matching in photo-z space, but 
photo-z estimates are subject to bias. 
Therefore we replace the redshift binning by an equivalent
aperture in the photometric observables. 

\subsection{Matching Distributions: Observables} 

For concreteness we take the photometric observables to be the 
$N_m$ magnitudes of each galaxy, where $N_m$ is the number of 
filter passbands in the survey: the magnitude vector of 
the $\alpha^{\rm th}$ galaxy in a sample is 
${\bmath m}_\alpha=m_\alpha^a$, with $a=1,...,N_m$. 
This choice of observables is not unique: we could instead use colors,
morphological information, or any other photometric observable.
A cell in magnitude space of radius $d_m$  
defines an $N_m$-dimensional hypervolume, $V_m=d_m^{N_m}$. 
The magnitude density in multi-magnitude space at point ${\bmath m}$
within $V_m$ is defined as
\begin{eqnarray}
\rho({\bmath m}) \equiv \frac{N({\bmath m})}{V_m} \,,
\label{eqn:den_m_gen_def}
\end{eqnarray}

\noindent where $N({\bmath m})$ is the number of objects in the
corresponding magnitude region.

The redshift distribution of the photometric set can be rewritten as
\begin{eqnarray}
P(z^i)^{\rm P} = \frac{\rho(z^i)^{\rm P}}{N_{\rm tot}^{\rm P}}
               = \int d{\bmath m} P(z^i|{\bmath m})^{\rm P} \frac{\rho({\bmath m})^{\rm P}}{N_{\rm tot}^{\rm P}}\,. 
\end{eqnarray}

Similarly, the distribution of the weighted training is given by
\begin{eqnarray}
P(z^i)^{\rm T}_{\rm wei} = \rho(z^i)^{T} W_{\alpha}
                         = \int d{\bmath m} P(z^i|{\bmath m})^{\rm T} \rho({\bmath m})^{\rm T} W_{\alpha}\,. 
\end{eqnarray}

Motivated by Eq.~(\ref{eqn:wei_z_gen_def}) and given our desire to set 
$P(z^i)^{\rm T}_{\rm wei}=P(z^i)^{\rm P}$, we
redefine the galaxy weights -- now as a function of magnitude 
densities -- as
\begin{eqnarray}
W_{\alpha} = \frac{1}{N_{\rm tot}^{\rm P}}\frac{\rho({\bmath m})^{\rm P}}{\rho({\bmath m})^{\rm T}} \,.
\label{eqn:wei_m_gen_def}
\end{eqnarray}

Therefore, the weighted training set distribution will provide an
unbiased estimate of the true distribution in the photometric set if
\begin{eqnarray}
\int d{\bmath m} P(z^i|{\bmath m})^{\rm P} \rho({\bmath m})^{\rm P}                          
            &=& \int d{\bmath m} P(z^i|{\bmath m})^{\rm T} \rho({\bmath m})^{\rm P} \,.
\label{eqn:wei_cond}
\end{eqnarray}

Notice that for a given magnitude ${\bmath m}$, there could be
a broad range of possible redshifts $z$ due to degeneracies, 
and the weighting method would still work as long as 
Eq.~(\ref{eqn:wei_cond}) is satisfied.
One obvious instance where this happens is if
\begin{equation}
P(z^i|{\bmath m})^{\rm P} = P(z^i|{\bmath m})^{\rm T}, 
\label{eqn:wei_assumption}
\end{equation}

\noindent i.e. if the training and photometric sets have the exact 
same degeneracies between redshift and magnitudes.
A training set may violate this condition 
if it has selection effects that are very different from those 
of the photometric set. 
The selection effect can easily be accounted for if it happens 
in the space of observables.
However, effects due to spectroscopic failure and large-scale
structure (LSS) are more difficult to model and control. 

\begin{figure}
    \begin{center}
      \resizebox{80mm}{!}{\includegraphics[angle=-90]{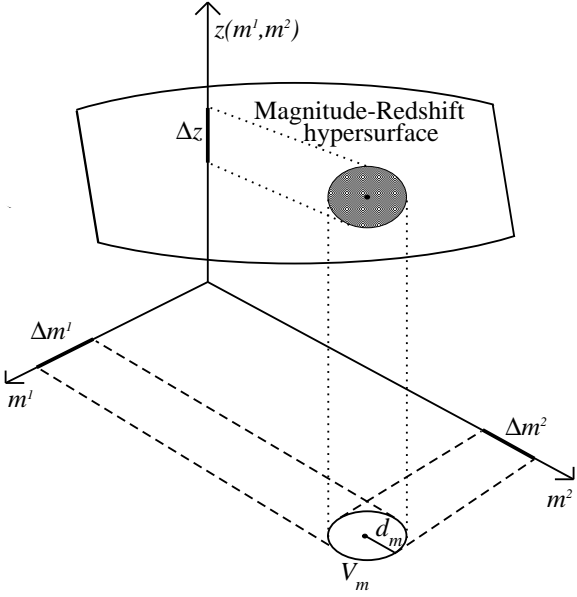}}
    \end{center}
\caption{Idealized magnitude-redshift hypersurface for $N_m=2$ magnitudes. 
Without degeneracies, the hypervolume in magnitude space surrounding a galaxy, 
 $V_{m} \propto d_m^{N_m}$, corresponds to an 
approximate redshift interval $\Delta z$. 
Whereas empirical photo-z 
methods usually make this implicit assumption, the weighting method works
under more general conditions.
}\label{fig_zdist}
\end{figure}

In the case where degeneracies are small,
$P(z^i|{\bmath m})$ approaches a delta function 
$\delta(z^i|{\bmath m}) $ 
and we have $\rho(z_i)=\rho({\bmath m})$, i.e. the magnitude hypervolume $V_m$
specifies uniquely a corresponding cell in redshift 
$\Delta z$ as indicated in  Fig.~\ref{fig_zdist}. 
The latter is typically the assumption of empirical photo-z methods;
violation of this condition leads to photo-z biases and spurious
peaks in the photo-z distribution.
In contrast, the weighting method works more
generally, since the only requirement is that the redshift distribution
inside $V_m$ must be the same for training and photometric sets.

In order to calculate weights for training set galaxies using 
Eq.~(\ref{eqn:wei_m_gen_def}), we estimate 
the density $\rho({\bmath m})$ using the nearest-neighbor prescription 
described below.

\subsection{Neighbors in Magnitude Space}

A nearest-neighbor approach to calculating the density of 
galaxies in 
magnitude space is advantageous,
because it enables control of statistical errors (shot noise) 
while also ensuring adequate ``locality'' of
the cells in magnitude space. 
The distance $d_{\alpha\beta}$ in magnitude space 
between the $\alpha^{\rm th}$ and $\beta^{\rm th}$ galaxies in a 
(photometric or spectroscopic) sample is defined by 
\begin{eqnarray}
(d_{\alpha\beta})^2 \equiv ({\bmath m}_\alpha-{\bmath m}_\beta)^2
                       =\sum_{a=1}^{N_m} (m_{\beta}^{a}-m_{\alpha}^{a})^2 \,.
\label{eqn:dist_gen_def}
\end{eqnarray}

\noindent We use this distance to find the set of 
{\it nearest neighbors} to the $\alpha^{\rm th}$ object, i.e., 
the set of galaxies with the smallest $d_{\alpha\beta}$. 
The density in magnitude space around this object, $\rho({\bmath m}_\alpha)$, 
is then estimated as the ratio of the
number of nearest neighbors $N_{\rm nei}$ to the magnitude hypervolume 
$V_m$ that they 
occupy, cf. Eq.~(\ref{eqn:den_m_gen_def}). 
For fixed 
$N_{\rm nei}$, if we order the neighbors by their 
distance from the $\alpha^{\rm th}$ galaxy, then we can define the hypervolume 
by the distance from $\alpha$ to the $(N_{\rm nei})^{\rm th}$ (most distant) 
neighbor, indexed by $\gamma$, $V_m = (d_{\alpha\gamma})^{N_m}$. 

Estimating the local 
density in the spectroscopic training set 
using a fixed (non-zero) value for $N({\bmath m}_\alpha)^{\rm T}=N_{\rm nei}$ 
ensures that 
the density estimate is positive-definite and that the resulting  
weight is well defined. To estimate the corresponding density in the 
photometric 
sample, we simply count the number of galaxies in the photometric sample, 
$N({\bmath m}_\alpha)^{\rm P}$, that occupy the {\it same} hypervolume $V_m$ 
around the point ${\bmath m}_\alpha$. 
Since the densities are estimated in the spectroscopic and photometric sets 
using the same hypervolume, the ratio of the densities is 
simply the ratio of the corresponding numbers of objects within the volume,  
and the weight for the $\alpha^{\rm th}$ training-set galaxy is 
therefore given by
\begin{eqnarray}
W_{\alpha} =  \frac{1}{N_{\rm tot}^{\rm P}} 
              \frac{ N({\bmath m}_\alpha)^{\rm P} }{ N({\bmath m}_\alpha)^{\rm T} } \,.
\label{eqn:wei_def_num} 
\end{eqnarray}

The optimal choice of $N_{\rm nei}$ balances locality against 
statistical errors. By locality we mean that the 
distance to the $(N_{\rm nei})^{\rm th}$ nearest neighbor, $d_{\alpha\gamma}$,  
should ideally be smaller than the characteristic scale in magnitude 
space over which $\rho({\bmath m})$ varies; this argues for small hypervolumes, 
i.e., small values of $N_{\rm nei}$. On the other hand, if $N_{\rm nei}$ 
is chosen too small, the resulting estimate of 
the density, $\rho({\bmath m})=N_{\rm nei}/V_m$, 
will suffer from large shot-noise error. The resulting statistical 
error on the weight is 
\begin{eqnarray}
\frac{\delta W_{\alpha}}{W_{\alpha}} 
                     = \left[ \frac{1}{N({\bmath m}_\alpha)^{\rm P}} 
                            + \frac{1}{N({\bmath m}_\alpha)^{\rm T}}\right]^{1/2} \,.
\end{eqnarray}

The optimal value of $N_{\rm nei}$
will depend on the characteristics of the photometric and spectroscopic 
samples at hand and should be determined using mock catalogs. 
For the DES and SDSS catalogs, we find in \S~\ref{sec:res} that 
the quality of the $N(z)$ reconstruction is relatively insensitive to 
the choice of $N_{\rm nei}$. 
The results we present there use 
the optimal values of $N_{\rm nei}$ determined by trial and error.

This implementation of the nearest-neighbor approach to estimating the 
magnitude-space densities and weights is not unique. 
For example, we could have instead used the same number of neighbors in 
both the spectroscopic and photometric samples, in which case the 
weights would be given by the ratio of corresponding hypervolumes,
\begin{eqnarray}
W_{\alpha} = \frac{1}{N_{\rm tot}^{\rm P}} \left [ 
\frac{(d_{\alpha\gamma})^{\rm T} }
{ (d_{\alpha\gamma'})^{\rm P} } \right]^{N_m} \, ,
\end{eqnarray}

\noindent where $\gamma'$ indicates the $(N_{\rm nei})^{\rm th}$ 
nearest neighbor in the photometric sample. 
Our tests indicate that this produces similar results 
to the fixed hypervolume method, but that the latter is slightly more 
stable in sparsely occupied regions of the spectroscopic and photometric 
samples. In a region of magnitude space that is sparsely occupied 
in the photometric sample, using a fixed number of objects can 
result in a non-local estimate of the density. 
Fixing the hypervolume instead tends to avoid that problem.
The results we present in \S~\ref{sec:res} 
use the fixed hypervolume, Eq.~(\ref{eqn:wei_def_num}), to estimate the 
weights.

\subsection{Weight Renormalization} \label{subsec:ren}

    If the spectroscopic training set has significantly different 
distributions of photometric observables than the photometric 
sample, then there may be galaxies in the training set that 
have very few or no neighbors in the photometric sample. Such galaxies 
will receive very small or zero weight and therefore make 
no contribution to the estimate of $N(z)^{\rm P}$. In this case, 
a recalculation of
the weights may improve the accuracy of the redshift distribution 
reconstruction.

    The idea is to perform a recalculation similar to a renormalization 
procedure. After an initial calculation of the weights, we
remove objects from the training set that were assigned very small or 
zero weights. Using the objects that remain, the weights are 
recalculated, possibly using a smaller number of neighbors to achieve more
locality of the new weights. This procedure can be iterated until some
convergence of the weights is achieved.

    As this renormalization procedure is iterated, the 
distribution in photometric observables of the remaining 
training set objects will approach that of 
the photometric sample, and the weights become more homogeneous. 

We have found the renormalization to be useful 
if a large fraction of the training set objects have
very small or zero weights. 
However, we do not expect to apply renormalization in
practical situations, since the training set will typically be
much smaller than the photometric set of interest.
We suggest the use of simulations to study if
the renormalization may help or not in each case. 
We present a case in which the renormalization significantly improves
the weighting in \S~\ref{sec:res}.

\subsection{Summary of the Algorithm}
To summarize, we outline the steps of the algorithm 
used to estimate the redshift distribution $N(z)$ of a photometric 
sample:

\begin{itemize}
\item  For each galaxy $\alpha$ in the spectroscopic training set, 
find a fixed number $N({\bmath m}_\alpha)^{\rm T}=N_{\rm nei}$ of its  
nearest neighbors in the training set according to the distance defined in 
Eq.~(\ref{eqn:dist_gen_def}) 
and compute the cell radius  $d_{\alpha\gamma}$ as the distance to the 
$(N_{\rm nei})^{\rm th}$ nearest neighbor.
\item Find the number $N({\bmath m}_\alpha)^{\rm P}$ 
of objects in the photometric sample that fall within the same cell 
radius (volume).

\item Compute the weight $W_{\alpha}$ according to Eq.~(\ref{eqn:wei_def_num}).

\item Repeat the weight calculation for each galaxy in the spectroscopic 
training set. Estimate the redshift distribution $P(z^i)^{\rm P}$ by 
summing the weights for all training-set galaxies in the $i^{\rm th}$ 
redshift bin, cf. Eq.~(\ref{eqn:p_wei_def}).

\item If a large number of training-set galaxies have very low or zero 
weight, the renormalization procedure of \S~\ref{subsec:ren} can 
be implemented.

\end{itemize}

\subsection{Weighting vs. Photo-z's}\label{subsec:WvsPz}

It is worth contrasting the key assumption of the weighting method--that 
samples with identical distributions of photometric observables 
have identical distributions of redshift--with 
the stronger assumption implicit in training-set based photo-z estimates. 
Photo-z estimators assume that there is (and try to find) a functional 
correspondence between a set of photometric observables 
and redshift; 
degeneracies in that correspondence lead to photo-z biases. 
For the weighting method, all that is assumed is that values of the  
photometric observables uniquely determine the redshift {\it probability 
distribution} of galaxies with those observables, a distribution  
which may be multiply peaked, as long as these features appear in 
both the training and photometric sets. 
Moreover, the weights from a number of 
training-set galaxies are summed to estimate $N(z^i)$ in a given 
redshift bin. If that number is reasonably large, this stacking  will
tend to cancel out possible statistical errors in individual galaxy weights. 
In Cunha et al. (in preparation),  we show how the weighting procedure 
can be used to estimate a redshift distribution $p(z)$ for each galaxy 
in the photometric sample and thereby avoid the biases of photo-z estimates.
\cite{Manetal07} show 
that using this weighted $p(z)$ in place of  
photo-z's is very effective in  
reducing calibration biases in galaxy-galaxy weak lensing. 

In \S~\ref{sec:res}, we will present results for both the weighting method 
and photo-z estimates for comparison.

\begin{figure*}
  \begin{center}
    \begin{minipage}[t]{160mm}
      \resizebox{160mm}{!}{\includegraphics[angle=0]{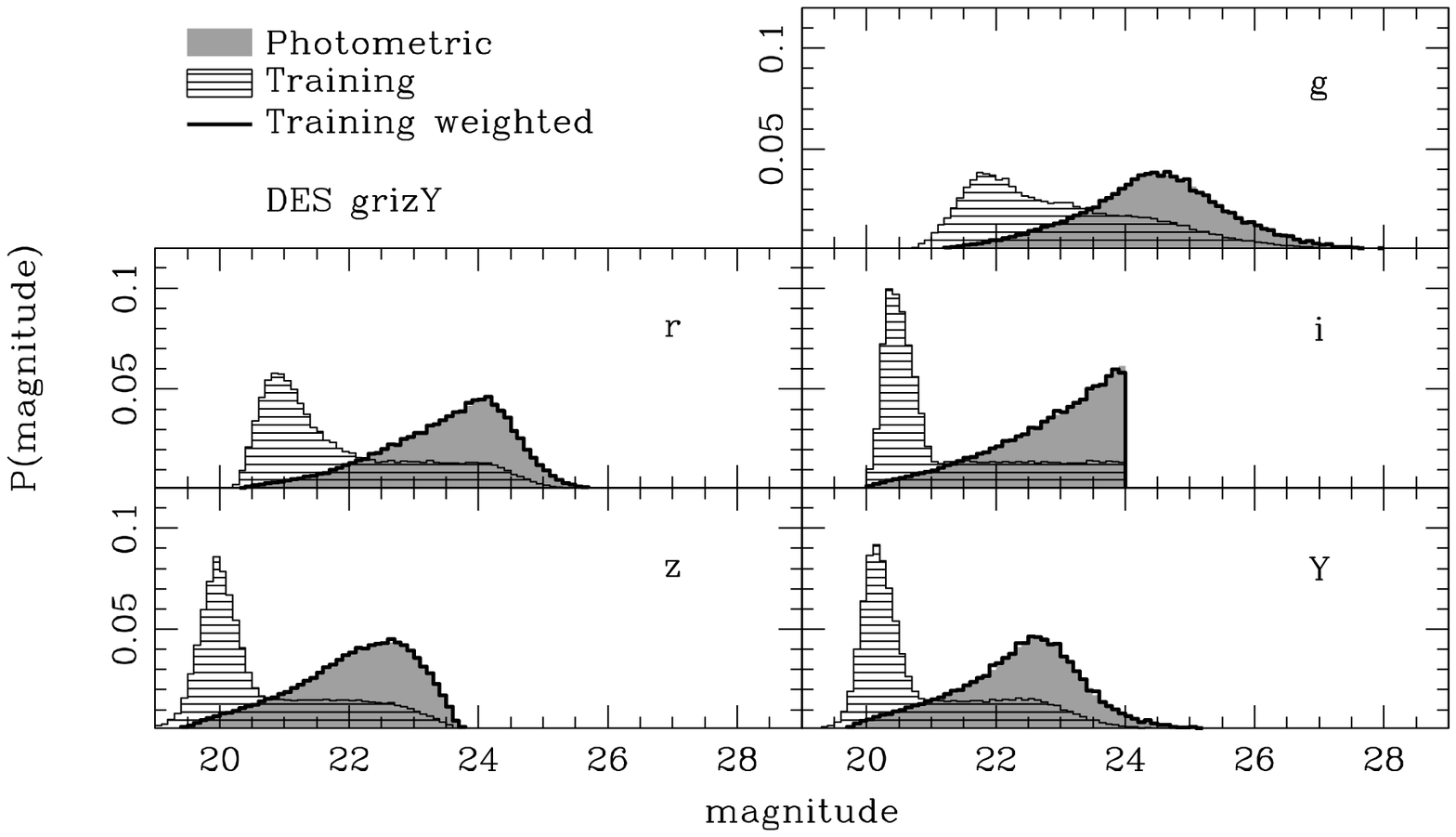}}
    \hfill
      \resizebox{160mm}{!}{\includegraphics[angle=0]{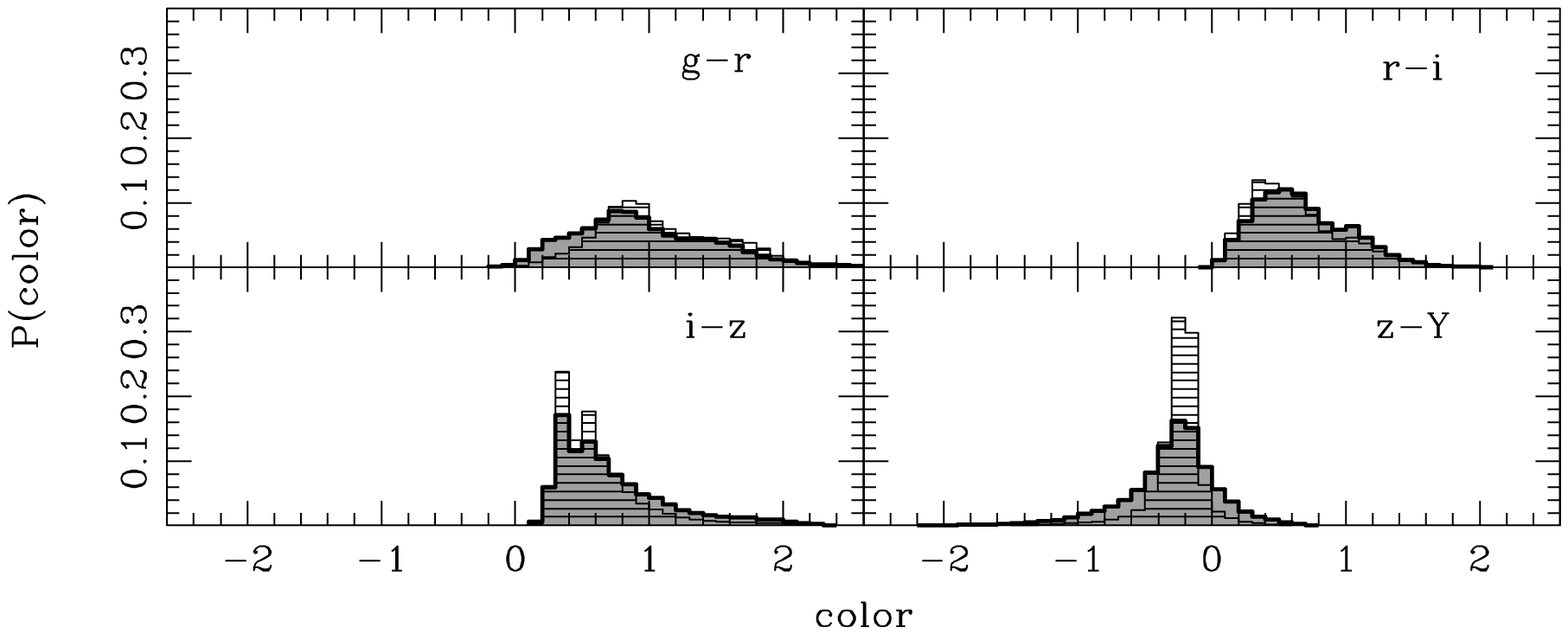}}
    \end{minipage}
  \end{center}
\caption{Distributions of magnitudes $grizY$ and colors 
$g-r$, $r-i$, $i-z$, $z-Y$, 
for the DES mock photometric  
catalog and for the first spectroscopic training set. 
Grey regions indicate the distributions 
in the photometric sample, horizontal hatched regions indicate those 
for the spectroscopic training set, and the solid black histograms 
are those for the weighted training set. 
}\label{fig:all.mag.col.dist.DESmock.tflat.igfdf}
\end{figure*}

\section{Measures of Reconstruction Quality} \label{sec:qua}
We measure the quality of the estimated redshift distribution 
reconstruction using two simple metrics.
The first is the $\chi^2$ statistic (per degree of freedom and 
per galaxy), defined as
\begin{eqnarray}
(\chi^{2})^{\rm X}&=&\frac{1}{N_{\rm bin}-1}\sum_{i=1}^{N_{\rm bin}}
            \frac{\left[P(z^{i})^{\rm X}-P(z^{i})^{\rm P}\right]^2}
                 {P(z^{i})^{\rm P}} \,, 
\end{eqnarray}

\noindent where $N_{\rm bin}$ is the number of redshift bins used, and 
$P(z^{i})^{\rm X}$ is equal to 
$P(z^{i})^{\rm T}_{\rm wei}$ if the weighting procedure is used 
or to 
$P(z_{\rm phot}^{i})^{\rm P}$ if the redshift distribution is instead 
estimated using photo-z's. The usual definition of $\chi^2$
uses the numbers $N(z^i)$ of objects in given bins instead
of the normalized probability $P(z^i)$; multiplying our
$\chi^2$ by $\Delta z N_{\rm tot}$ gives the usual definition. 
We chose the above version 
so the resulting quantity is independent of the number of galaxies 
and the number of redshift bins.
The definition allows us to more fairly compare reconstruction qualities
across different data sets.
Because the probabilities are
normalized, the number of degrees of freedom is $N_{\rm bin}-1$.

The second metric we employ 
is the Kolmogorov-Smirnov (KS) statistic, defined as 
the maximum difference between the two cumulative redshift distributions 
being compared, for example, the cumulative distributions corresponding 
to  $P(z^{i})^{\rm T}_{\rm wei}\Delta z$ and 
$P(z^{i})^{\rm P} \Delta z$. 
The KS statistic is more sensitive to the changes in the median of the two 
compared distributions whereas the $\chi^2$ tends to stress 
regions of the distribution that are least well sampled, i.e. regions
where $P(z^i)$ is small.
In our implementation, we use binned cumulative distributions instead of
unbinned cumulative distributions, and therefore our metric is not strictly
the KS statistic.

It is important to stress that we do not associate any fundamental 
meaning to the absolute values of the metrics introduced above.
They are used solely to compare the qualities of different 
reconstructions relative to each other.

For the DES mock catalogs, we 
use $N_{\rm bin}=50$ redshift bins covering the 
redshift interval $z=0-2$. For the real catalogs based on
SDSS photometry, we use 
$N_{\rm bin}=30$ bins over $z=0 - 1.2$. 
In both sets of catalogs, the bins are equally sized 
in redshift.

\section{Results} \label{sec:res}
In this section, we test the methods of reconstruction 
of the redshift distribution on several simulated and real data sets. 

\subsection{DES mock catalog}  \label{subsec:DESres}

\begin{figure*}
  \begin{center}
    \begin{minipage}[t]{160mm}
      \resizebox{80mm}{!}{\includegraphics[angle=0]{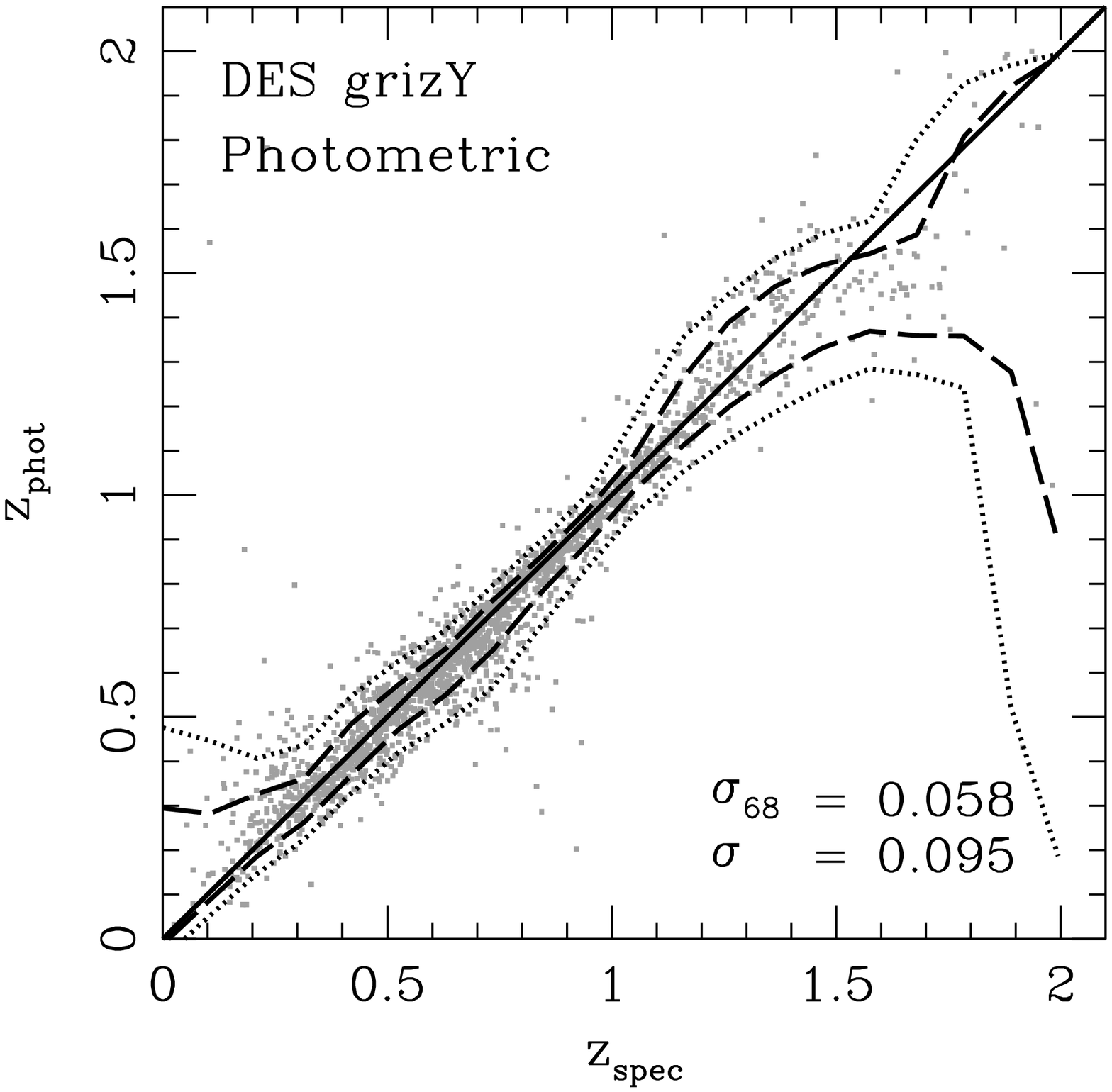}}
      \hfill
      \resizebox{80mm}{!}{\includegraphics[angle=0]{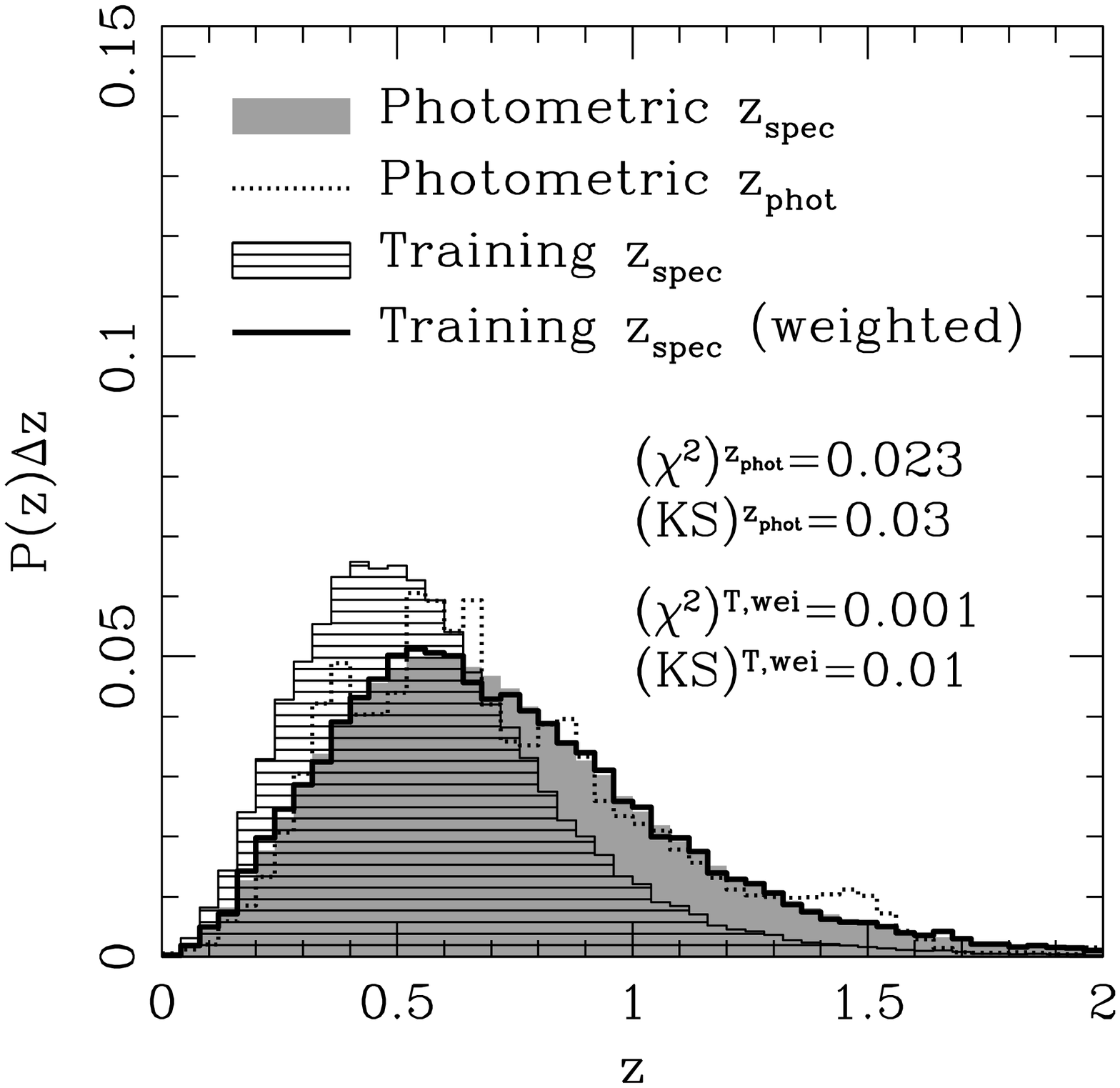}}
    \end{minipage}
  \end{center}
  \caption{
    {\it Left panel:} 
    Photometric redshift $z_{\rm phot}$ vs. 
    spectroscopic redshift $z_{\rm spec}$  
    for a random sampling of 
    galaxies in the DES mock catalog. 
    Photo-z's were computed using the neural network algorithm described 
    in Appendix A, using the first spectroscopic training set described in 
    the text. 
    The dashed and dotted curves are the contours containing 
    $68\%$ and $95\%$ of the galaxies in narrow bins of $z_{\rm spec}$.
    Also indicated are the overall rms photo-z scatter $\sigma$ 
    and $68\%$ confidence region $\sigma_{68}$ (see their definition in the text). 
    {\it Right panel:} Redshift distributions. 
    The shaded grey region 
    shows the redshift distribution of the photometric sample that 
    we are aiming to reconstruct. 
    The horizontal hatched distribution shows the redshift 
    distribution of the spectroscopic training set corresponding to the 
    magnitude and color distributions shown in 
    Fig.~\ref{fig:all.mag.col.dist.DESmock.tflat.igfdf}. 
    The solid black histogram shows the reconstructed redshift distribution 
    using the weighting method. 
    The dotted lines show the neural network photo-z 
    distribution of the photometric set, showing peaks due to 
    photo-z biases.
    Also indicated are the $\chi^2$ and KS statistics for both the 
    weighting method and the photo-z distribution. 
  }\label{fig:all.z.dist.DESmock.tflat.igfdf}
\end{figure*}

We first consider the DES mock photometric catalog of 500,000 galaxies  
described in \S~\ref{subsec:DESmock}. We test the reconstruction 
using two spectroscopic training sets comprising $100,000$ galaxies each.
Each of them have different distributions of magnitude, redshift, 
and galaxy type from the photometric sample. 

For the first training set, the magnitude and type 
distributions $P(i)$ and $P(t)$ differ from those of the 
photometric sample (recall the latter are given by 
Eq.~(\ref{eqn:Pi}) and Fig.~\ref{fig:prob.DESmock.fid}), but the 
conditional redshift probability 
$P(z|i)$ is identical to that of the photometric sample, Eq.~\ref{eqn:Pzi}. 
In particular, the spectroscopic 
$i$-magnitude distribution, shown as the horizontal 
hatched region in Fig.~\ref{fig:all.mag.col.dist.DESmock.tflat.igfdf}, 
is skewed toward brighter magnitudes (and therefore lower redshift) 
than that of the photometric 
sample, with a peak at $i \simeq 20.5$, 
though it does include galaxies to the photometric 
limit, $i=24$. The spectroscopic type distribution $P(t)$ is 
chosen to be flat over the interval $t=-0.5$ to 3.5, in contrast 
to the bimodal 
photometric type distribution shown in Fig.~\ref{fig:prob.DESmock.fid}.
The spectroscopic type distribution may differ from that of the 
photometric sample due to, e.g., color selection in spectroscopic 
targeting or higher spectroscopic efficiency for certain galaxy types. 

The weights were computed from Eq.~\ref{eqn:wei_def_num}, using the
nearest neighbors in a fixed hypervolume in the space of all colors 
and $i$ magnitude. 
The hypervolume for each training set galaxy 
was defined by the $N_{\rm nei}=16$ 
nearest neighbors in the training set. Our tests indicated that 
this value of $N_{\rm nei}$ yields the lowest value for $\chi^2$ 
for the reconstruction in this case. However, as noted earlier, 
the results are not sensitive to this choice. Increasing 
$N_{\rm nei}$ to 64 causes a negligible change in $\chi^2$, while 
decreasing it to 4 causes an increase of less than 20\% in $\chi^2$. 

Figs.~\ref{fig:all.mag.col.dist.DESmock.tflat.igfdf} and 
\ref{fig:all.z.dist.DESmock.tflat.igfdf} display 
the reconstruction results for this case using the weighting 
method. Fig.~\ref{fig:all.mag.col.dist.DESmock.tflat.igfdf} shows 
the distributions of magnitudes and colors 
for the photometric sample (solid grey), 
the spectroscopic training sample (horizontal hatched), and 
the weighted training set (black line). The coincidence of the 
grey and black regions demonstrates that the weighted training 
set distributions in magnitudes and colors are excellent matches to those  
of the photometric sample. The right panel of 
Fig.~\ref{fig:all.z.dist.DESmock.tflat.igfdf} 
shows that the weighted training-set redshift distribution also 
provides a precise estimate of the redshift distribution of the 
photometric sample. The measures of reconstruction quality for this match are 
$(\chi^2)^{\rm T,wei}=0.001$ and (KS)$^{\rm T,wei}=0.01$.

\begin{figure*}
  \begin{center}
   \begin{minipage}[t]{160mm}
      \resizebox{80mm}{!}{\includegraphics[angle=0]{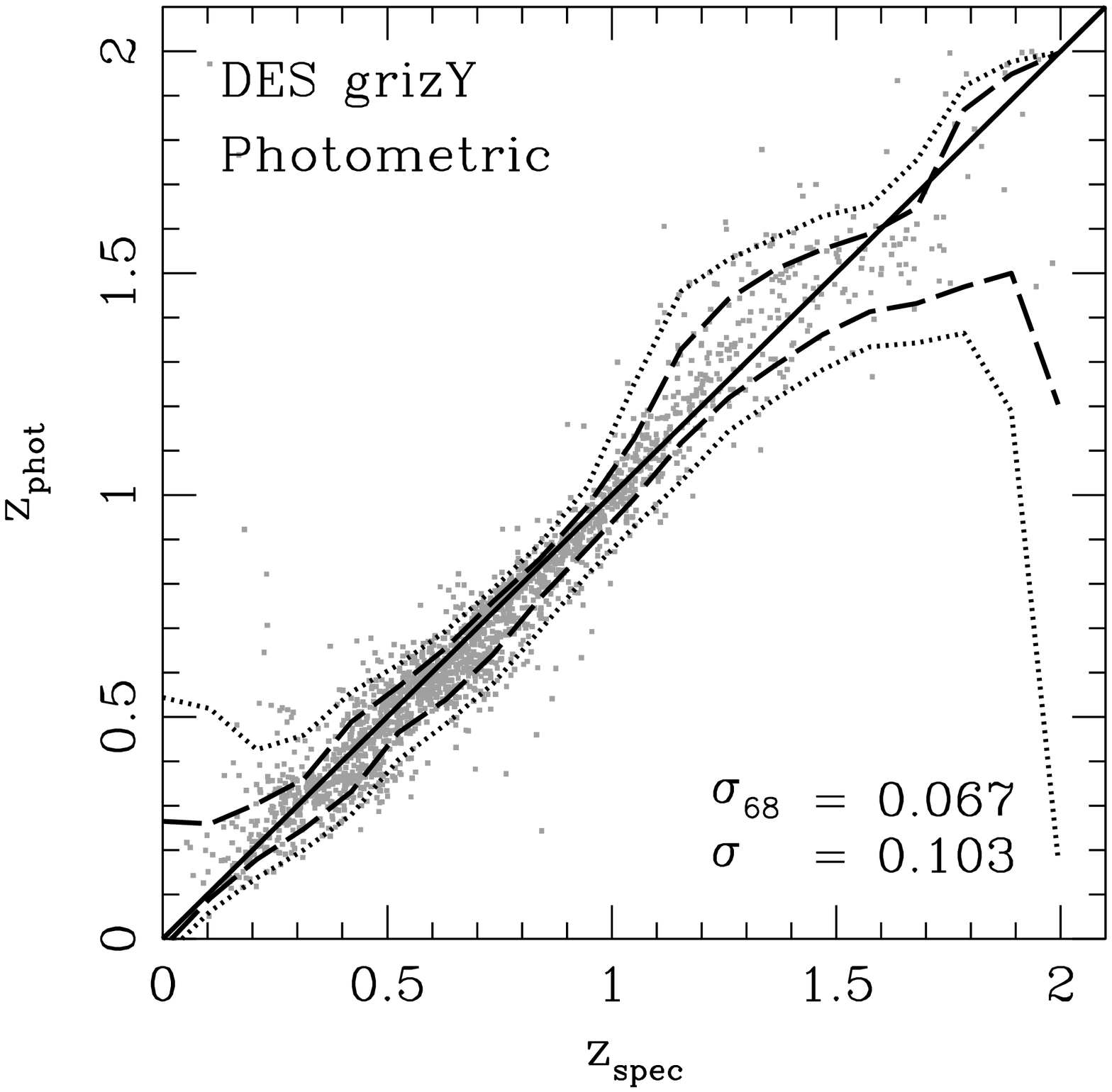}}
    \hfill
      \resizebox{80mm}{!}{\includegraphics[angle=0]{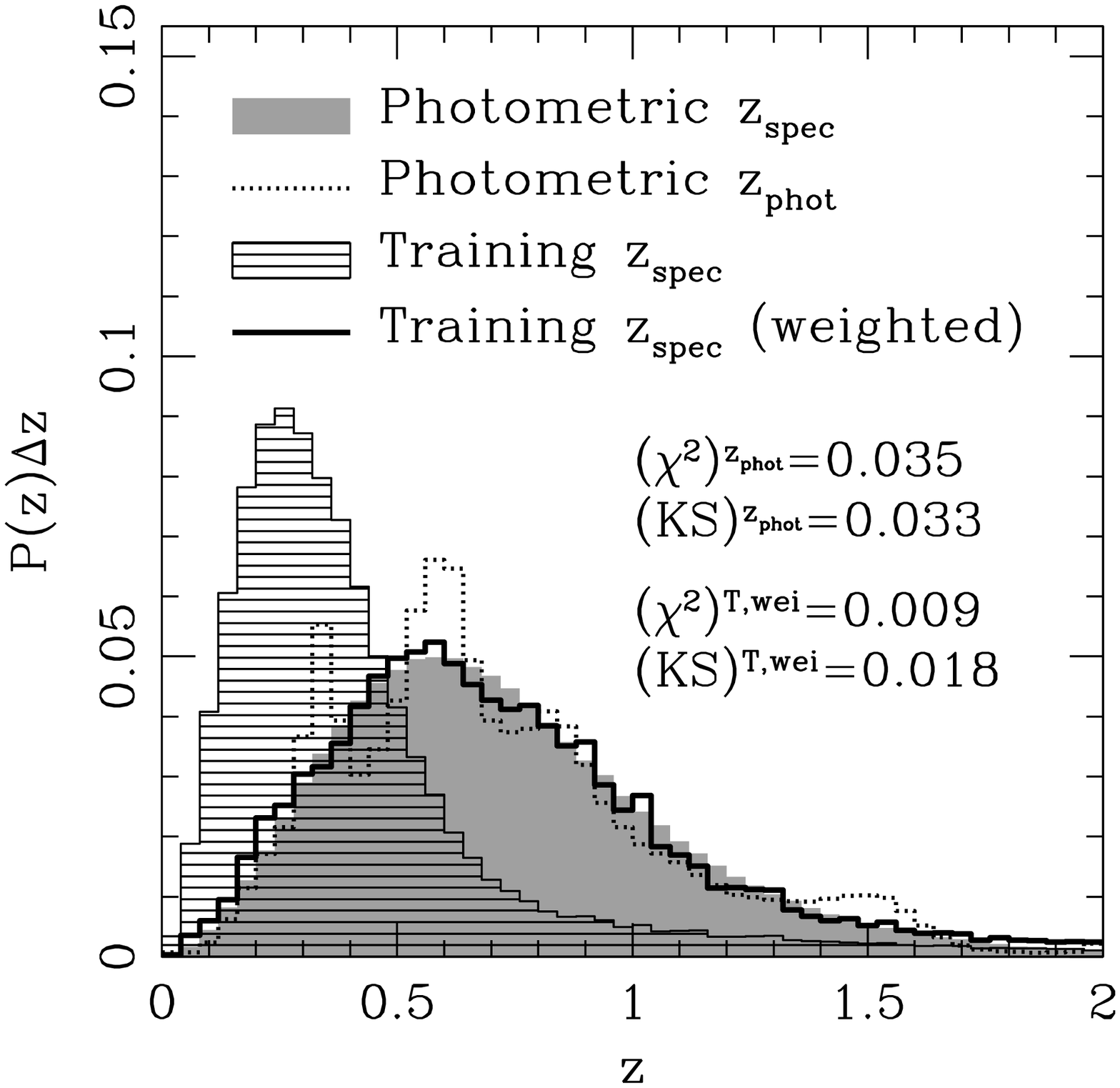}}
    \end{minipage}
  \end{center}
\caption{As Fig. \ref{fig:all.z.dist.DESmock.tflat.igfdf}, but now 
using the second spectroscopic training set described in the text. 
In this case, the training set is even less representative of the 
photometric sample than in the previous example, but the weighting 
procedure still accurately reconstructs the redshift distribution 
of the photometric sample.
}\label{fig:z.dist.DESmock.zstanmod.tflat.igfdf}
\end{figure*}

For comparison, we also carry out the $N(z)$ reconstruction using 
photometric redshift estimates. 
The left panel of 
Fig.~\ref{fig:all.z.dist.DESmock.tflat.igfdf} shows the photo-z scatter 
for the neural network photo-z estimator described in Appendix A. 
Here the spectroscopic set has been split into 2 samples 
(the training and validation sets) of
equal sizes that were used to train 
and validate the network, which was finally applied to compute 
photo-z's for the photometric sample. 
To test the overall quality of the photo-z 
estimates we use two photo-z performance metrics, whose values
are also displayed in 
Fig.~\ref{fig:all.z.dist.DESmock.tflat.igfdf}. 
The first metric is the photo-z {\it rms} 
scatter, $\sigma$, averaged over all $N$ objects in the photometric 
set, defined by 
\begin{eqnarray}
\sigma^2&=&\frac{1}{N}\sum_{i=1}^{N}\left(z_{\rm phot}^{i}-z_{\rm spec}^{i}\right)^2 ~,
\end{eqnarray}

\noindent whereas, the second performance metric, 
denoted by $\sigma_{68}$, is 
the range containing $68\%$ of the photometric set objects in the 
distribution of $\delta z = z_{\rm phot}-z_{\rm spec}$. 
We also define similarly $\sigma_{68}$ and $\sigma_{95}$ in bins 
of $z_{\rm spec}$, and the dashed and dotted lines in the left
panel of 
Fig.~\ref{fig:all.z.dist.DESmock.tflat.igfdf} show these regions
respectively.

The right panel of 
Fig.~\ref{fig:all.z.dist.DESmock.tflat.igfdf} shows the resulting 
$N(z_{\rm phot})$ distribution for the photometric sample (dotted line). 
Due to degeneracies in the relation between magnitudes and 
redshift, the 
photo-z estimate is biased at low and high redshifts. 
In particular, 
the photo-z solution produces an excess of galaxies at 
$z_{\rm phot} \approx 0.4$, 0.6, and $1.4$, which  
translates into the 
spurious peaks at these redshifts in the right panel of 
Fig.~\ref{fig:all.z.dist.DESmock.tflat.igfdf}. The corresponding 
measures of reconstruction quality are 
$(\chi^2)^{z_{\rm phot}}=0.023$ and (KS)$^{z_{\rm phot}}=0.03$, 
significantly worse than those for the weighting method. 
Deconvolution of the $N(z_{\rm phot})$ distribution can improve 
this match \citep{pad05}; 
we will explore that elsewhere (Cunha et al., in preparation).

For the second training set example, we make the spectroscopic 
sample even less representative of the photometric sample. We keep 
the spectroscopic 
$i$-magnitude and type distributions of the previous example, but 
we alter the conditional redshift probability $P(z|i)$ 
of the training set so that it is more 
concentrated toward lower redshift while still covering the redshift
range $z \in [0,2]$.  
Specifically, we change the parameter values that determine 
$z_d$ and $\sigma_d$ in Eqs.~(\ref{eqn:zd}) and (\ref{eqn:sigd}) 
to $(b_1,b_2,b_3)=(-0.5,0.6,-0.5)$ and
$(c_1,c_2,c_3)=(0.38,0.02,3.5)$. By 
decreasing the values of $b_i$ relative to those of the photometric 
sample, we shift the distribution toward 
lower redshift, while increasing $c_2$ and $c_3$ increases the 
spread of the distribution in redshift so that the full range to 
$z=2$ is still covered. 
This example could correspond, for instance, to a training set that is 
obtained by combining different spectroscopic surveys with different
selection functions. 
Notice that changing only $P(z|i)$ means that we are
changing only one dimension of the probability $P(z|{\bmath m})$
which lives in a 5-dimensional space of magnitudes.
If there are no further selection effects, we still expect 
the weighting method to work reasonably well, though obviously
not as accurately as in the first case of 
Fig.~(\ref{fig:all.z.dist.DESmock.tflat.igfdf}).

In this case, we find that $N_{\rm nei}=4$ neighbors is nearly optimal:
training sets that are less representative require fewer 
neighbors to provide the best match, since locality in magnitude/color 
space becomes more important. 
The redshift distribution of the photometric sample estimated from 
the weighted training set is shown in the right panel of 
Fig.~\ref{fig:z.dist.DESmock.zstanmod.tflat.igfdf}. (As in the 
previous example, 
the weighted reconstructions of the magnitude and color distributions 
are nearly perfect, as in Fig.~\ref{fig:all.mag.col.dist.DESmock.tflat.igfdf}, 
so we do not show them.) 
Even though the training-set redshift distribution is now considerably 
different from that of the photometric sample, peaking at $z \sim 0.25$ 
as opposed to $z \sim 0.6$, the weighting method 
still does a very good job of estimation, with 
$(\chi^2)^{\rm T,wei}=0.009$ and (KS)$^{\rm T,wei}=0.018$.  

The left panel of 
Fig.~\ref{fig:z.dist.DESmock.zstanmod.tflat.igfdf} shows the scatter plot 
for the neural network photo-z estimates for this training set; the 
photo-z scatter is larger than in the previous example, 
as expected since the training set is 
less representative of the photometric sample. As the right panel of 
Fig.~\ref{fig:z.dist.DESmock.zstanmod.tflat.igfdf} also shows, 
the photo-z distribution has spurious peaks at the same redshifts 
as before, but they are now more pronounced. 
The photo-z distribution has
$(\chi^2)^{z_{\rm phot}}=0.035$ and (KS)$^{z_{\rm phot}}=0.033$,
significantly worse than for the weighting method.

\subsection{SDSS Data Catalogs} \label{subsec:SDSSres}

\begin{figure*}
  \begin{center}
    \begin{minipage}[t]{160mm}
      \resizebox{160mm}{!}{\includegraphics[angle=0]{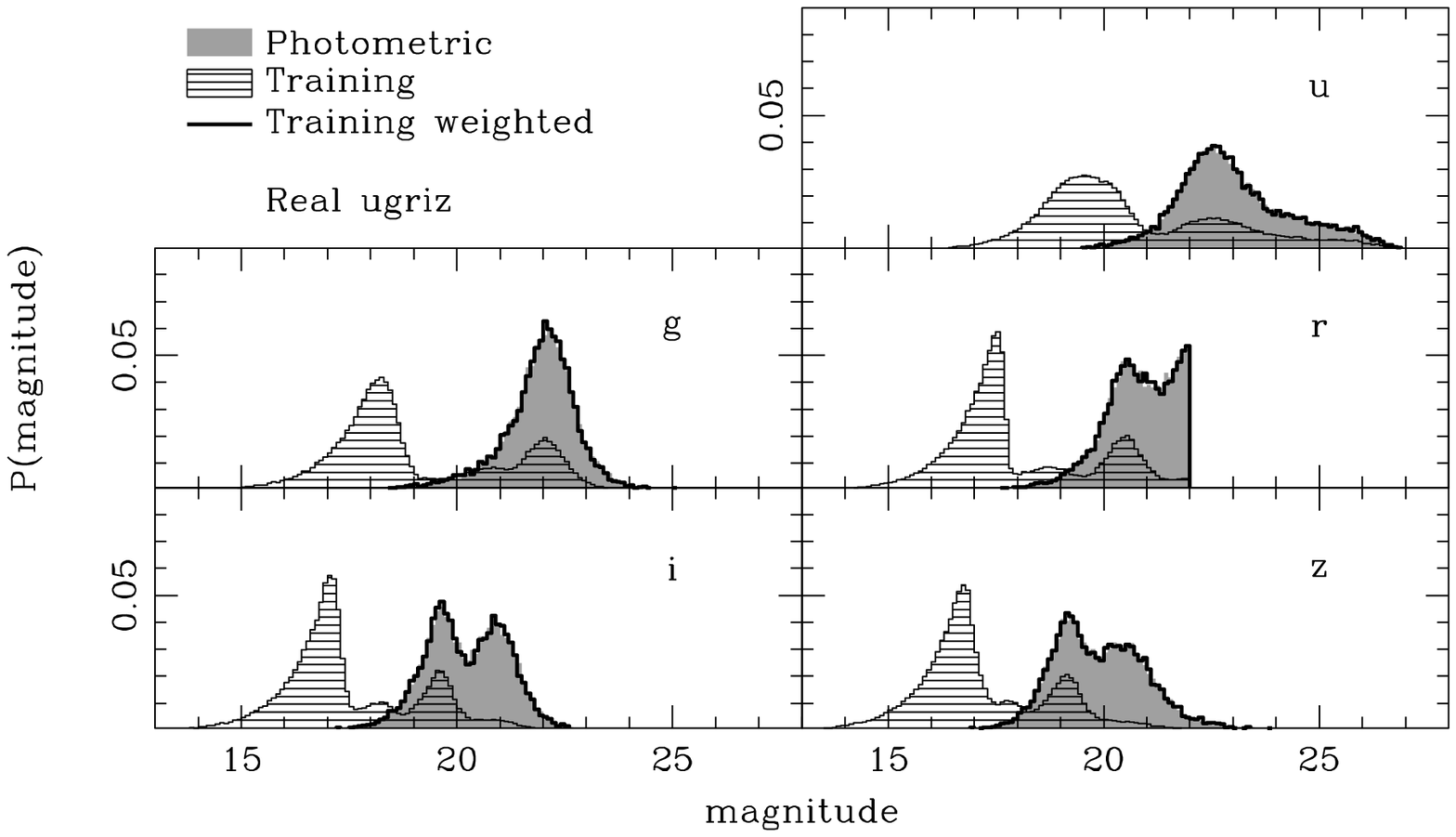}}
  \hfill
      \resizebox{160mm}{!}{\includegraphics[angle=0]{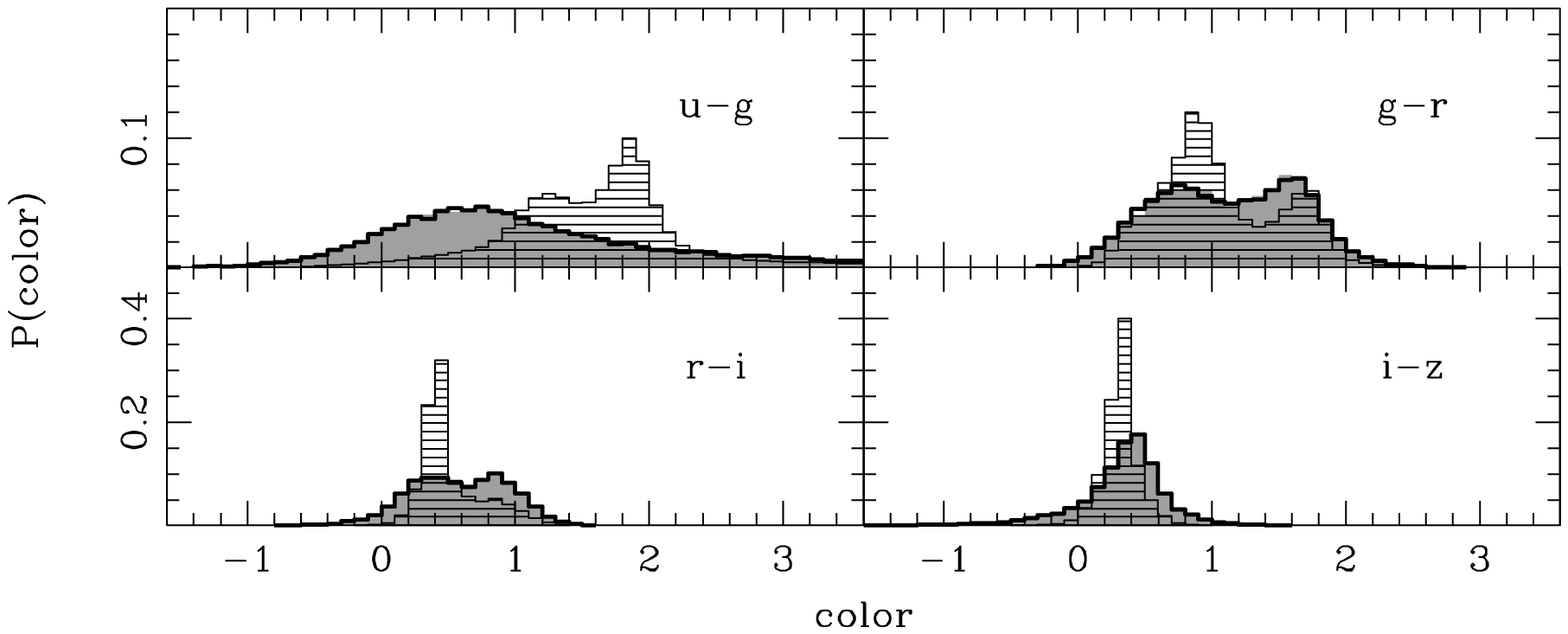}}
    \end{minipage}
  \end{center}
\caption{Distributions of magnitudes ($ugriz$) and colors 
($u-g$, $g-r$, $r-i$, $i-z$) for samples drawn from SDSS DR6 photometry, 
for the first example in the text in \S~\ref{subsec:SDSSres}. 
Grey regions denote the distributions in the photometric sample, 
horizontal hatched regions are for the spectroscopic training set, 
and the black histograms show the reconstructed distributions for 
the photometric sample using the weighted training set.
}\label{fig:all.mag.col.dist.real.photomix}
\end{figure*}

\begin{figure*}
  \begin{center}
   \begin{minipage}[t]{160mm}
      \resizebox{80mm}{!}{\includegraphics[angle=0]{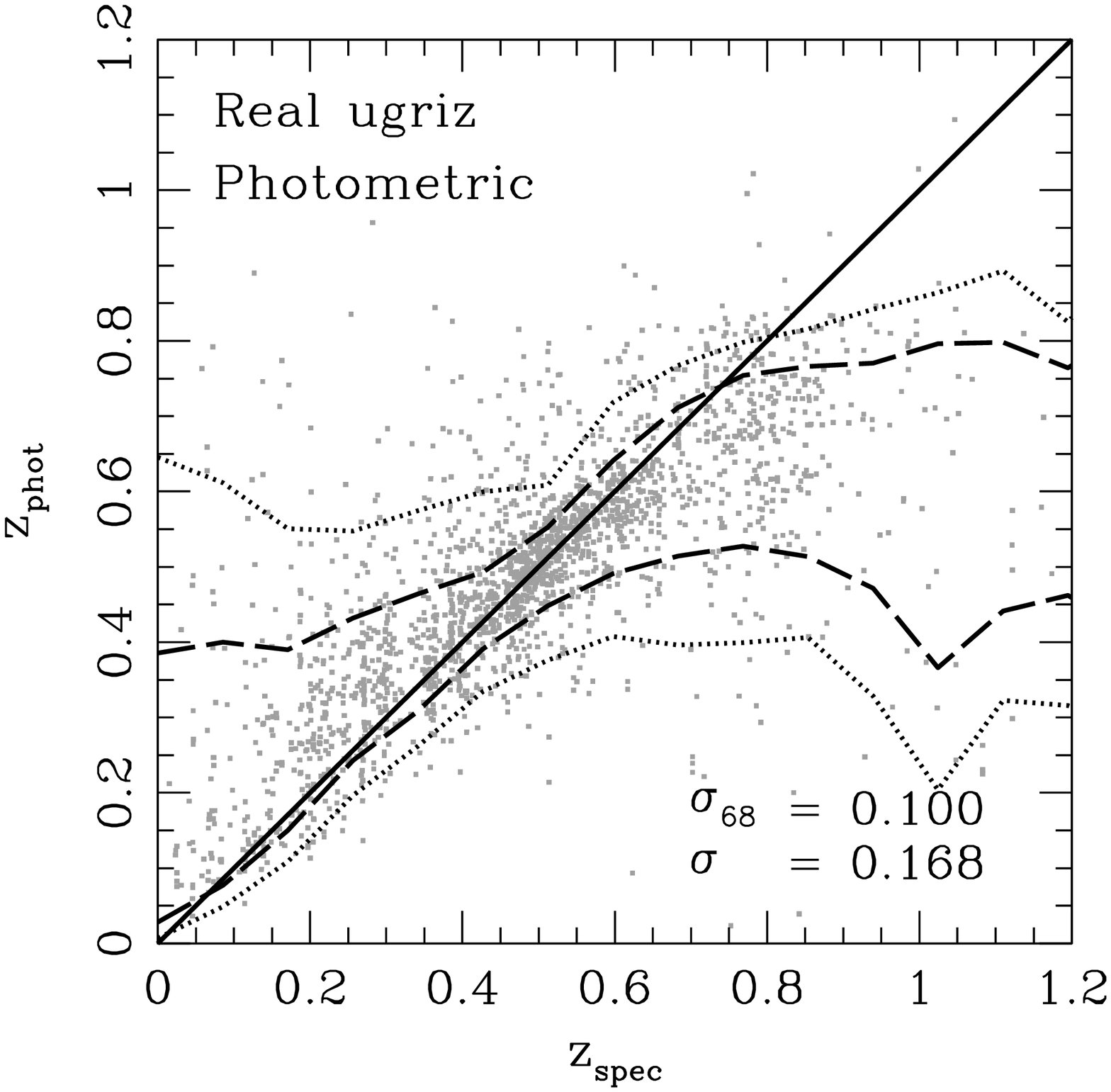}}
    \hfill
      \resizebox{80mm}{!}{\includegraphics[angle=0]{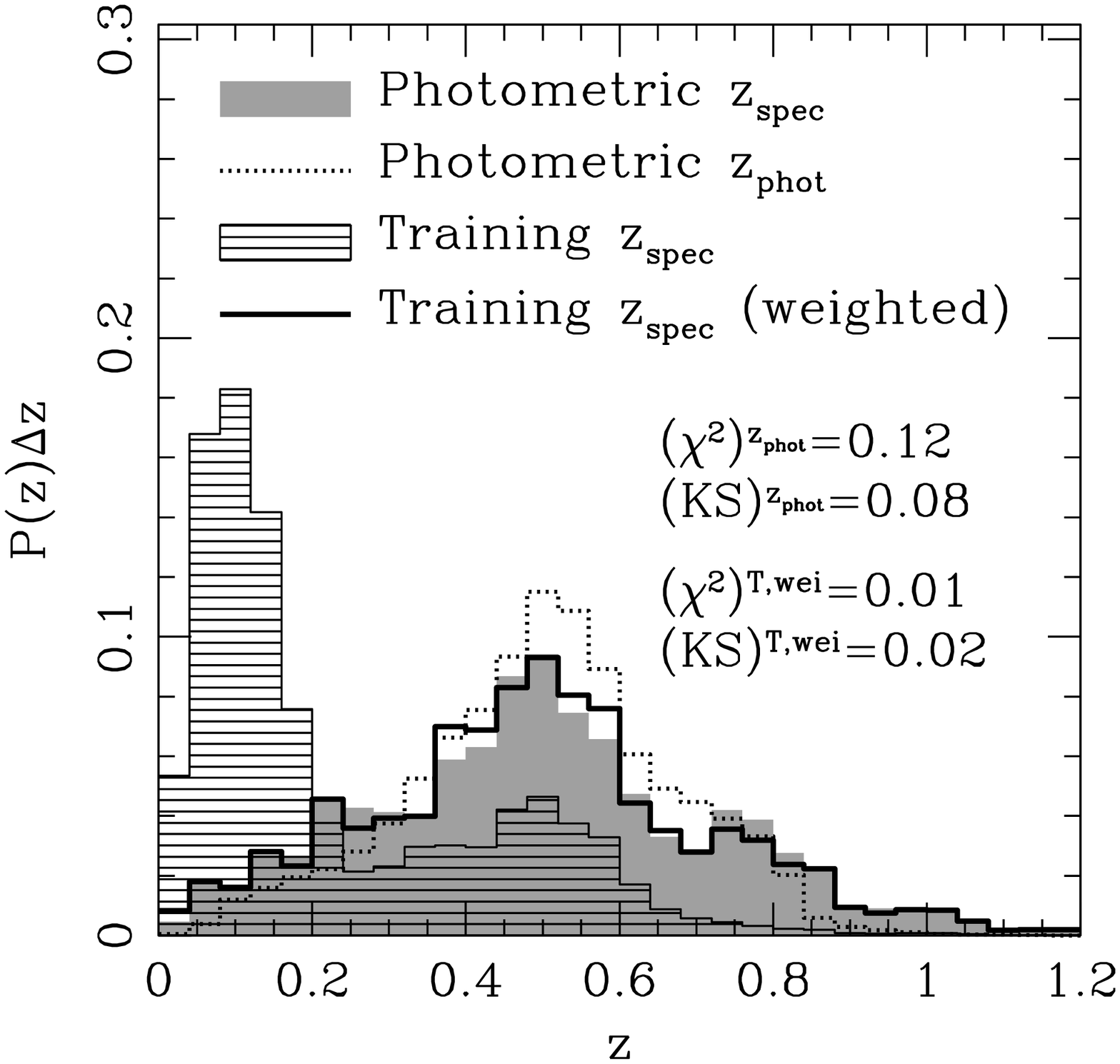}}
    \end{minipage}
  \end{center}
\caption{{\it Left panel:} Scatter of neural network photo-z estimates 
using the same training set and photometric sample as in Fig. 
\ref{fig:all.mag.col.dist.real.photomix}. {\it Right panel:} 
Redshift distributions for the photometric sample, spectroscopic training set, 
weighted training set, and photo-z's.  
}\label{fig:z.dist.real.photomix}
\end{figure*}

Here we consider two examples of 
the reconstruction of the redshift distribution 
for photometric samples drawn from the SDSS, using the 
spectroscopic samples described in \S~\ref{subsec:SDSScat} and 
shown in Fig.~\ref{fig:dndrz_real}. 

For the first case, we created a spectroscopic training set comprising 
$200,000$ galaxies from the SDSS spectroscopic survey,  
$15,000$ galaxies from CNOC2, 
$6,000$ from the DEEP+DEEP2 sample, and
$47,000$ from 2SLAQ,
for a total of $268,000$ galaxies. 
For all these
sets, the galaxies were randomly selected from the parent spectroscopic 
sample. 
The photometric sample comprises the remaining galaxies with 
spectroscopic redshifts, namely 
$5,381$ galaxies from CNOC2, 
$1,541$ from CFRS, 
$5,040$ from DEEP+DEEP2, 
$2,078$ from DEEP2/EGS, 
$654$ from TKRS, and
$5,762$ from 2SLAQ,
for a total of $20,456$ galaxies. 

For this example, we calculated weights for the training-set galaxies 
using a hypervolume in color/$r$-magnitude space with $N_{\rm nei}=32$ 
neighbors. 
Fig.~\ref{fig:all.mag.col.dist.real.photomix} 
shows the magnitude and color distributions for the photometric sample, 
the spectroscopic training set, and the weighted training set. 
The weighting procedure provides an excellent match to the 
distributions for the photometric sample. This is not a difficult 
test for the method since, with the exception of the SDSS 
spectroscopic sample, the distributions for the training and 
photometric samples are rather similar and by construction
Eq.~(\ref{eqn:wei_assumption}) is satisfied. 
The right panel of  
Fig.~\ref{fig:z.dist.real.photomix} shows the corresponding  
redshift distributions for these training and photometric samples. 
The weighted training set provides a good estimate of $N(z)$ for 
the photometric sample, 
with $(\chi^2)^{\rm T,wei}=0.01$ and (KS)$^{\rm T,wei}=0.02$. 
The left panel of Fig.~\ref{fig:z.dist.real.photomix} shows the 
photo-z scatter for the neural network photo-z estimator trained on 
the same spectroscopic sample and applied to the photometric sample.   
The biases at low and high redshift are evident. The photo-z distribution 
shown in the right panel of Fig.~\ref{fig:z.dist.real.photomix} provides 
a less accurate representation of the true redshift distribution of 
the photometric sample than the weighting procedure--features in 
the true redshift distribution are smoothed out, and the distribution 
is systematically underestimated at high redshift; the 
corresponding $(\chi^2)^{z_{\rm phot}} =0.12$ and 
(KS)$^{z_{\rm phot}}=0.08$ are again considerably worse than for
the weighting procedure.

For the second case, the training set and the photometric sample 
come from different spectroscopic surveys.  
Here, the training set comprises the galaxies from all the spectroscopic 
surveys with the exception of the DEEP2/EGS catalog, and the latter 
is taken to be the photometric sample. The training set 
contains $286,378$ galaxies, and  
the photometric sample $2078$. 
Since DEEP2/EGS is -- apart from the match to SDSS photometry -- roughly flux limited, this provides a more realistic
case, except for the fact that the photometric sample in practice 
would typically be much larger.

In this case, since the training set is much larger than the
photometric sample, the 
best results are achieved if the weights are renormalized 
according to the procedure described in \S~\ref{subsec:ren}.
However let us first consider what happens if we do not apply
renormalization and compute the weights only once as in all
previous cases. 

Matching both colors and $r$-magnitude is better achieved with high
number of neighbors. The maximum number we chose was  
$N_{\rm nei}=4096$, in which case we obtain $(\chi^2)^{\rm T,wei}=0.29$.
On the other hand, if we only perform the match in color space,
the best results happen with $N_{\rm nei}=1$ and 
also produce $(\chi^2)^{\rm T,wei}=0.29$. 
In the first case, we find that the redshift distribution 
is well reconstructed at low redshifts, but overestimated at higher
redshifts, whereas the opposite happens in the latter case of matching
only the color distributions.
These features suggest that we employ the following 
renormalization procedure described below.

\begin{figure*}
  \begin{center}
   \begin{minipage}[t]{160mm}
      \resizebox{80mm}{!}{\includegraphics[angle=0]{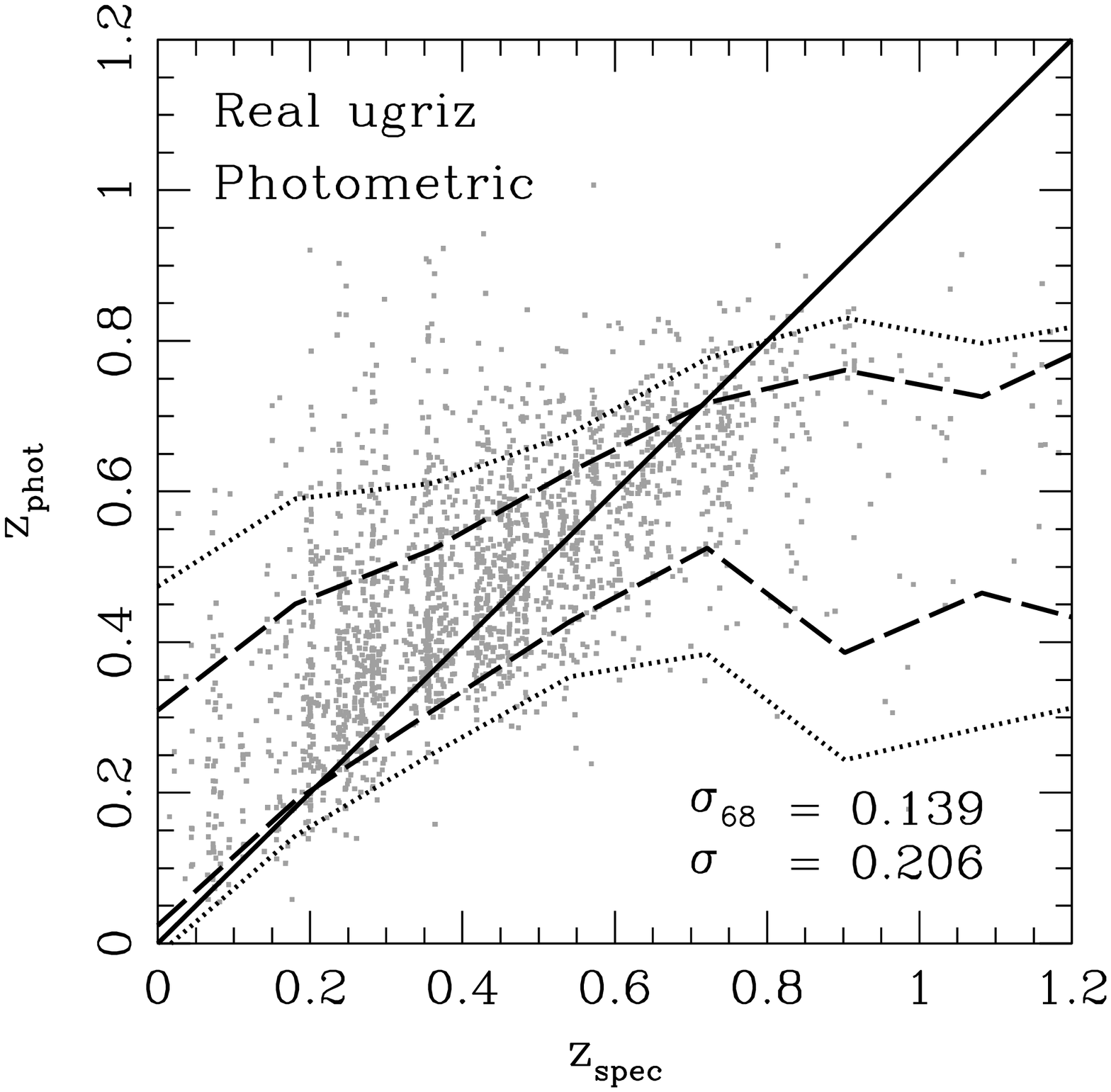}}
    \hfill
      \resizebox{80mm}{!}{\includegraphics[angle=0]{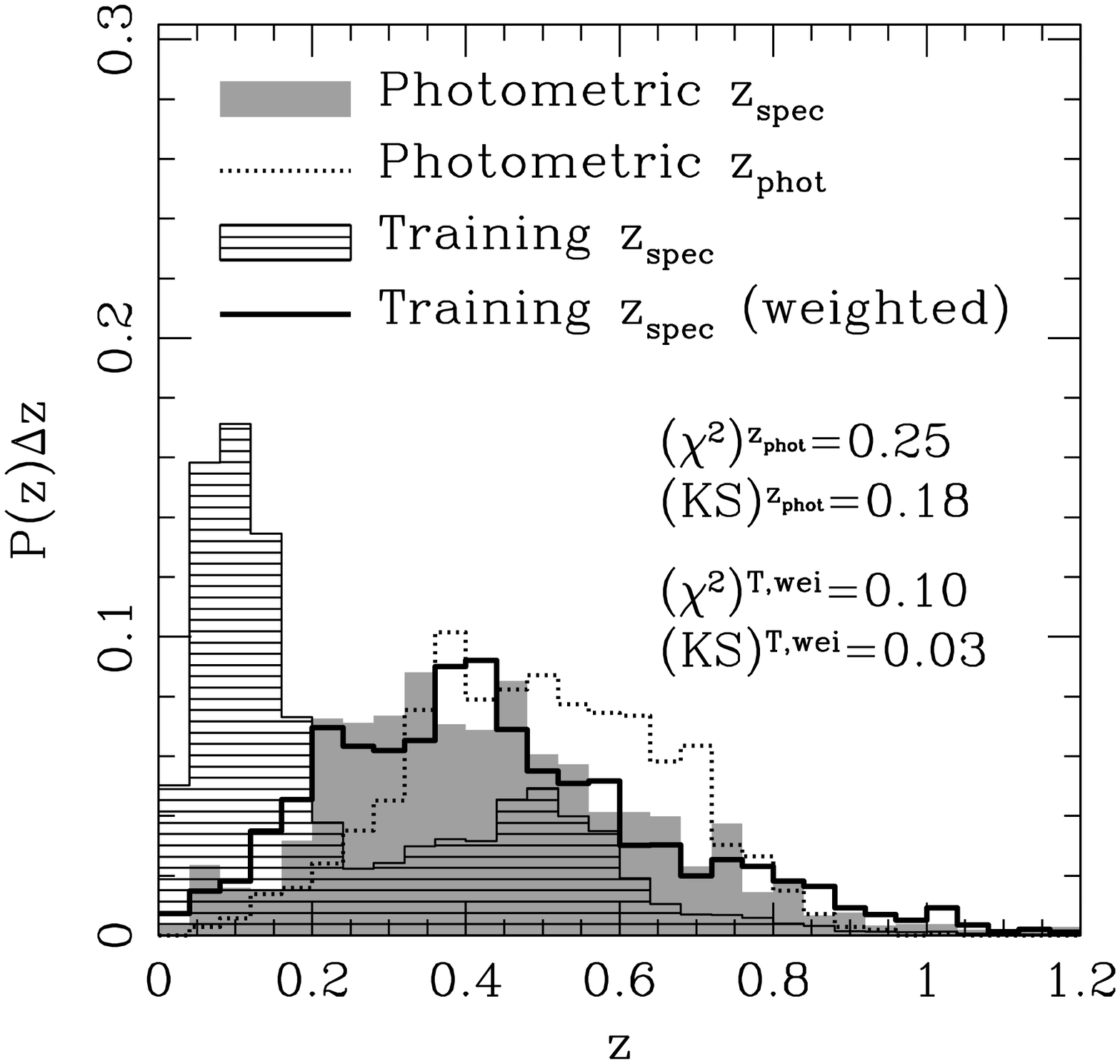}}
    \end{minipage}
  \end{center}
\caption{{\it Left panel:} Neural network photo-z scatter for the 
second case described in \S~\ref{subsec:SDSSres}, which uses 
DEEP2/EGS as the photometric sample and all other spectroscopic 
catalogs for the training set. {\it Right panel:} 
Redshift distributions for the photometric sample, training set, 
weighted training set, and photo-z. 
}\label{fig:z.dist.real.DEEP2-EGS}
\end{figure*}

We first calculate 
the weights by matching the distributions of colors and magnitudes using
$N_{\rm nei}=4096$ neighbors in the training set. 
After this first calculation, more than half of the training set
galaxies have zero weights and are removed from the catalog. 
We then iterate the weight calculation by matching only the color 
distributions. 
In each iteration, we remove objects with zero weight 
and reduce $N_{\rm nei}$ by a factor of 2 
until $N_{\rm nei}=1$. 
Only $7,968$ of the original training-set galaxies have positive weight 
in the final iteration. 
The right panel of 
Fig.~\ref{fig:z.dist.real.DEEP2-EGS} shows the resulting 
redshift distributions for this case; for the renormalized 
weighting procedure, the reconstruction has 
$(\chi^2)^{\rm T,wei}=0.10$, while the corresponding photo-z distribution has
$(\chi^2)^{\rm z_{\rm phot}}=0.25$; likewise
(KS)$^{\rm T,wei}=0.03$ whereas (KS)$^{z_{\rm phot}}=0.18$. 
It is clear that the weighting 
method provides a better estimate of $N(z)$. 

Given the small 
size of the photometric sample, there is considerable shot noise 
in its redshift distribution.
In addition, the small angular-area of the survey introduces significant
LSS effects in the redshift distribution \citep[see e.g.][]{Manetal07}.
The weighted reconstruction works well in spite of these complications.

Because we need to go down to $N_{\rm nei}=1$,
imposed by the locality requirement, Poisson errors of
individual galaxy weights are relatively large. 
However, given the large number of galaxies in each 
redshift bin, these errors cancel out and the overall reconstruction
is improved.

This case illustrates that this method has the potential
to provide very accurate estimations of the redshift distributions
of flux limited samples in future galaxy surveys, even when
they are subject to LSS effects.

\section{Discussion} \label{sec:dis}

We have presented a new 
technique to estimate the underlying redshift distribution
of photometric galaxy samples. The method relies on a spectroscopic 
training set and re-weighting of the training-set galaxies to match 
the distribution of photometric observables of the photometric sample. 
The weights are estimated using a flexible nearest-neighbor approach 
in color-magnitude space and the redshift distribution is estimated
by summing the galaxy weights in redshift bins. 
Tests on mock catalogs and on existing data sets show that this 
procedure yields an accurate estimate of the redshift distribution 
and that it performs significantly better than 
simply binning the photo-z estimates of individual galaxies 
in the photometric sample. 
The weighting method also appears to be robust, in the sense 
that the spectroscopic 
sample can have very different distributions of photometric 
observables and redshift from the photometric sample. 
The main requirement is that the training set should span the range 
of photometric observables found in the photometric sample.

The key assumption underlying the technique is that 
two samples (e.g., the spectroscopic and photometric) 
with the same distribution of photometric 
observables will have very similar redshift distributions. 
This assumption holds if the selection criteria used to define the two 
samples differ only in the space of photometric observables. 
Several effects can cause this condition to be violated: 
statistical errors, LSS and spectroscopic failures.

Statistical errors are the simplest to quantify and are significant in 
regions of magnitude space where the training 
set is sparse, typically at fainter magnitudes.
LSS can be significant if certain regions of the 
space of photometric observables are only represented in the training set 
by a survey that covers small solid angle, in which one or a few large 
structures dominate. 
We showed that, even in such cases, the weighting 
method works quite well (\S~\ref{subsec:SDSSres}).

Spectroscopic failures (i.e., targeted objects for which a redshift 
could not be obtained) in the training set can have a similar 
effect if the failures happen systematically, for instance in a particular 
galaxy spectral type. 
If the effects of spectroscopic failures are 
prevalent in regions of magnitude space where the redshift distribution 
is broad or multiply peaked, they can potentially cause systematic errors 
in the recovery of the redshift distribution.
However we also showed that the weighting method performs well even
when we take an arbitrary type distribution in the training set
(\S~\ref{subsec:DESres}).

The weighting method requires a training set with a size (density)
such that the inter-(training-set)-galaxy separation in the space of 
photometric observables is comparable to the characteristic (curvature) 
scale of the redshift/photometric-observables manifold or the scale 
defined by the typical photometric errors - whichever is smaller. 
That condition ensures that on average at least one neighbor to the galaxy 
is meaningful; in practice it would be safer to have the density a few 
times larger than this minimum density.
The use of mock catalogs can shed light on the optimal parameters
to employ on the weighting method, such as the number of 
neighbors $N_{\rm nei}$ (possibly varying according to the 
local density), the minimum training set size, the need
or not for renormalization, etc. 
Since these simulations are necessary for other typical calibration 
reasons, their need does not put any strong restrictions to the
application of the weighting method. 
For instance, simulations and calibration samples are necessary
to calibrate the photo-z errors. 
 
This weighting technique has been used to estimate the  
redshift distribution of 
the SDSS DR6 photometric 
sample~\citep{OyaLimCunLinFriShe08} and to 
help assess the quality of the 
photo-z's computed for that sample.
It has also been used in conjunction with photo-z's in the 
measurement of the SDSS 
cluster-mass cross-correlation function via 
weak lensing~\citep{Sheetal07a}, allowing for the inversion of 
cluster mass profiles~\citep{Johetal07} and estimation of cluster 
mass-richness relations~\citep{Johetal07} and
 mass-to-light ratios~\citep{Sheetal07b}.
Finally, this weighting scheme has recently been employed in the 
study of galaxy-galaxy 
weak lensing calibration bias~\citep{Manetal07}, 
where it was shown to yield much smaller biases than those 
arising from photo-z estimates. 
For future photometric surveys, the weighting method can 
complement and provide cross-checks on photo-z estimates and 
help control photo-z errors.

\section*{Acknowledgments}

We would like to thank Dinoj Surendran and Mark SubbaRao for useful 
discussions about nearest neighbor search methods and for introducing
the authors to a fast algorithm using Cover-Trees.
This work was supported by the KICP 
 under NSF No. PHY-0114422 and NSF PHY-0551142,
by NSF grants AST-0239759, AST-0507666, 
and AST-0708154 at the University of Chicago,
by the DOE at the University of Chicago and Fermilab, and by DOE 
contract number DE-AC02-07CH11359. 

Funding for the SDSS and SDSS-II has been provided by the Alfred P. 
Sloan Foundation, the Participating Institutions, the National Science 
Foundation, the U.S. Department of Energy, the National Aeronautics and
 Space Administration, the Japanese Monbukagakusho, the Max Planck 
Society, and the Higher Education Funding Council for England. The 
SDSS Web Site is {\tt http://www.sdss.org/}.

The SDSS is managed by the Astrophysical Research Consortium for the
 Participating Institutions. The Participating Institutions are the
 American Museum of Natural History, Astrophysical Institute Potsdam,
 University of Basel, University of Cambridge, Case Western Reserve
 University, University of Chicago, Drexel University, Fermilab, the
 Institute for Advanced Study, the Japan Participation Group, Johns
 Hopkins University, the Joint Institute for Nuclear Astrophysics,
 the Kavli Institute for Particle Astrophysics and Cosmology, the 
Korean Scientist Group, the Chinese Academy of Sciences (LAMOST), 
Los Alamos National Laboratory, the Max-Planck-Institute for Astronomy 
(MPIA), the Max-Planck-Institute for Astrophysics (MPA), New Mexico 
State University, Ohio State University, University of Pittsburgh, 
University of Portsmouth, Princeton University, the United States 
Naval Observatory, and the University of Washington.  

Funding for the DEEP2 survey has been provided by NSF grants AST95-09298, 
AST-0071048, AST-0071198, AST-0507428, and AST-0507483 as well as 
NASA LTSA grant NNG04GC89G.

Some of the data presented herein were obtained at the W. M. Keck 
Observatory, which is operated as a scientific partnership among the 
California Institute of Technology, the University of California and the 
National Aeronautics and Space Administration. 
The Observatory was made possible by the generous financial support of 
the W. M. Keck Foundation. 
The DEEP2 team and Keck Observatory acknowledge the very significant 
cultural role and reverence that the summit of Mauna Kea has always had 
within the indigenous Hawaiian community and appreciate the opportunity to 
conduct observations from this mountain.

\bibliographystyle{mn2e}
\bibliography{zdistribution}

\begin{thebibliography}{}

\bibitem[\protect\citeauthoryear{{Banerji}, {Abdalla}, {Lahav} \&
  {Lin}}{{Banerji} et~al.}{2007}]{Ban07}
{Banerji} M.,  {Abdalla} F.~B.,  {Lahav} O.,    {Lin} H.,  2007, ArXiv
  e-prints, 711

\bibitem[\protect\citeauthoryear{{Bruzual} \& {Charlot}}{{Bruzual} \&
  {Charlot}}{1993}]{bru93}
{Bruzual} A.~G.,  {Charlot} S.,  1993, \apj, 405, 538

\bibitem[\protect\citeauthoryear{{Cannon} et~al.,}{{Cannon}
  et~al.}{2006}]{can06}
{Cannon} R.,  et~al., 2006, \mnras, 372, 425

\bibitem[\protect\citeauthoryear{{Capak} et~al.,}{{Capak}
  et~al.}{2004}]{cap04}
{Capak} P.,  et~al., 2004, \aj, 127, 180

\bibitem[\protect\citeauthoryear{{Coleman}, {Wu} \& {Weedman}}{{Coleman}
  et~al.}{1980}]{col80}
{Coleman} G.~D.,  {Wu} C.~C.,    {Weedman} D.~W.,  1980, \apjs, 43, 393

\bibitem[\protect\citeauthoryear{{Collister} \& {Lahav}}{{Collister} \&
  {Lahav}}{2004}]{ColLah04}
{Collister} A.~A.,  {Lahav} O.,  2004, \pasp, 116, 345

\bibitem[\protect\citeauthoryear{{Cowie}, {Barger}, {Hu}, {Capak} \&
  {Songaila}}{{Cowie} et~al.}{2004}]{cow04}
{Cowie} L.~L.,  {Barger} A.~J.,  {Hu} E.~M.,  {Capak} P.,    {Songaila} A.,
  2004, \aj, 127, 3137

\bibitem[\protect\citeauthoryear{{Davis} et~al.,}{{Davis}
  et~al.}{2007}]{dav07}
{Davis} M.,  et~al., 2007, \apjl, 660, L1

\bibitem[\protect\citeauthoryear{{Davis}, {Newman}, {Faber} \&
  {Phillips}}{{Davis} et~al.}{2001}]{deep2}
{Davis} M.,  {Newman} J.~A.,  {Faber} S.~M.,    {Phillips} A.~C.,  2001, in
  {Cristiani} S.,  {Renzini} A.,   {Williams} R.~E.,  eds, Deep Fields {The
  DEEP2 Redshift Survey}.
pp 241--+

\bibitem[\protect\citeauthoryear{Huterer, Kim, Krauss \& Broderick}{Huterer
  et~al.}{2004}]{HutKimKraBro04}
Huterer D.,  Kim A.,  Krauss L.~M.,    Broderick T.,  2004, Astrophys. J., 615,
  595

\bibitem[\protect\citeauthoryear{Huterer, Takada, Bernstein \& Jain}{Huterer
  et~al.}{2006}]{HutTakBerJai06}
Huterer D.,  Takada M.,  Bernstein G.,    Jain B.,  2006, MNRAS, 366, 101

\bibitem[\protect\citeauthoryear{{Jain}, {Connolly} \& {Takada}}{{Jain}
  et~al.}{2007}]{jain07}
{Jain} B.,  {Connolly} A.,    {Takada} M.,  2007, Journal of Cosmology and
  Astro-Particle Physics, 3, 13

\bibitem[\protect\citeauthoryear{{Johnston} et~al.,}{{Johnston}
  et~al.}{2007}]{Johetal07}
{Johnston} D.~E.,  et~al., 2007, ArXiv e-prints, 709

\bibitem[\protect\citeauthoryear{{Lilly}, {Le Fevre}, {Crampton}, {Hammer} \&
  {Tresse}}{{Lilly} et~al.}{1995}]{lil95}
{Lilly} S.~J.,  {Le Fevre} O.,  {Crampton} D.,  {Hammer} F.,    {Tresse} L.,
  1995, \apj, 455, 50

\bibitem[\protect\citeauthoryear{{Lima} \& {Hu}}{{Lima} \&
  {Hu}}{2007}]{LimHu07}
{Lima} M.,  {Hu} W.,  2007, \prd, 76, 123013

\bibitem[\protect\citeauthoryear{{Lin} et~al.,}{{Lin}  et~al.}{1999}]{lin99}
{Lin} H.,  et~al., 1999, \apj, 518, 533

\bibitem[\protect\citeauthoryear{{Ma}, {Hu} \& {Huterer}}{{Ma}
  et~al.}{2006}]{MaHuHut06}
{Ma} Z.,  {Hu} W.,    {Huterer} D.,  2006, \apj, 636, 21

\bibitem[\protect\citeauthoryear{{Mandelbaum} et~al.,}{{Mandelbaum}
  et~al.}{2007}]{Manetal07}
{Mandelbaum} R.,  et~al., 2007, ArXiv e-prints, 709

\bibitem[\protect\citeauthoryear{{Oyaizu}, {Lima}, {Cunha}, {Lin}, {Frieman} \&
  {Sheldon}}{{Oyaizu} et~al.}{2008}]{OyaLimCunLinFriShe08}
{Oyaizu} H.,  {Lima} M.,  {Cunha} C.~E.,  {Lin} H.,  {Frieman} J.,    {Sheldon}
  E.~S.,  2008, ArXiv e-prints, 708

\bibitem[\protect\citeauthoryear{{Padmanabhan} et~al.,}{{Padmanabhan}
  et~al.}{2005}]{pad05}
{Padmanabhan} N.,  et~al., 2005, \mnras, 359, 237

\bibitem[\protect\citeauthoryear{{Poli} et~al.,}{{Poli}  et~al.}{2003}]{pol03}
{Poli} F.,  et~al., 2003, \apjl, 593, L1

\bibitem[\protect\citeauthoryear{{Sheldon} et~al.,}{{Sheldon}
  et~al.}{2004}]{she04}
{Sheldon} E.~S.,  et~al., 2004, \aj, 127, 2544

\bibitem[\protect\citeauthoryear{{Sheldon} et~al.,}{{Sheldon}
  et~al.}{2007a}]{Sheetal07a}
{Sheldon} E.~S.,  et~al., 2007a, ArXiv e-prints, 709

\bibitem[\protect\citeauthoryear{{Sheldon} et~al.,}{{Sheldon}
  et~al.}{2007b}]{Sheetal07b}
{Sheldon} E.~S.,  et~al., 2007b, ArXiv e-prints, 709

\bibitem[\protect\citeauthoryear{{Sheth}}{{Sheth}}{2007}]{Sheth07}
{Sheth} R.~K.,  2007, \mnras, 378, 709

\bibitem[\protect\citeauthoryear{{Weiner} et~al.,}{{Weiner}
  et~al.}{2005}]{wei05}
{Weiner} B.~J.,  et~al., 2005, \apj, 620, 595

\bibitem[\protect\citeauthoryear{{Wirth} et~al.,}{{Wirth}
  et~al.}{2004}]{wir04}
{Wirth} G.~D.,  et~al., 2004, \aj, 127, 3121

\bibitem[\protect\citeauthoryear{{Yee} et~al.,}{{Yee}  et~al.}{2000}]{yee00}
{Yee} H.~K.~C.,  et~al., 2000, \apjs, 129, 475

\bibitem[\protect\citeauthoryear{Zhan}{Zhan}{2006}]{Zha06}
Zhan H.,  2006, JCAP, 0608, 008

\bibitem[\protect\citeauthoryear{{Zhan} \& {Knox}}{{Zhan} \&
  {Knox}}{2006}]{ZhaKno06}
{Zhan} H.,  {Knox} L.,  2006, \apj, 644, 663

\end{thebibliography}

\appendix

\section{Artificial Neural Network Photo-z's} \label{app:neu}

For comparison with the weighting method, we use an 
Artificial Neural Network (ANN) method to estimate photometric redshifts \citep{ColLah04, 
OyaLimCunLinFriShe08}
We use a particular type of ANN called a Feed Forward Multilayer
Perceptron (FFMP), which 
consists of several nodes arranged in layers through which 
signals propagate sequentially. 
The first layer, called the input layer, receives the input photometric 
observables (magnitudes, colors, etc.). 
The next layers, denoted hidden layers, propagate signals until 
the output layer, whose outputs are the desired quantities, in this
case the photo-z estimate. 
Following the notation of \cite{ColLah04}, we denote a network with 
$k$ layers and $N_i$ nodes in the $i^{\rm th}$ layer as $N_1:N_2: ... :N_k$.

A given node can be specified by the layer it belongs to and the 
position it occupies in the layer. Consider a node in layer $i$ and 
position $\alpha$  with $\alpha=1,2,...,N_i$. 
This node, denoted $P_{i\alpha}$, receives
a total input $I_{i\alpha}$ and fires an output $O_{i\alpha}$ given by
\begin{eqnarray}
O_{i\alpha}=F(I_{i\alpha}) \,,
\end{eqnarray}  

\noindent where $F(x)$ is the activation function. 
The photometric observables are the inputs $I_{1\alpha}$ to the 
first layer nodes, which produce outputs $O_{1\alpha}$. 
The outputs $O_{i\alpha}$ in layer $i$ are
propagated to nodes in the next layer $(i+1)$, denoted $P_{(i+1)\beta}$,
with $\beta=1,2,..N_{i+1}$. 
The total input $I_{(i+1)\beta}$ is a weighted sum of the outputs 
$O_{i\alpha}$
\begin{eqnarray}
I_{(i+1)\beta} = \sum_{\alpha=1}^{N_i} w_{i\alpha\beta} O_{i\alpha},
\end{eqnarray}

\noindent where $w_{i\alpha\beta}$ is the weight that connects nodes 
$P_{i\alpha}$ and $P_{(i+1)\beta}$.
Iterating the process in layer $i+1$, signals propagate from hidden layer 
to hidden layer until the output layer.
There are various choices for the activation function $F(x)$ such as:
a sigmoid, a hyperbolic tangent, a step function, a linear function, etc.
This choice typically has no important effect
on the final photo-z's, and different activation functions can be used
in different layers. Training the network consists in finding
weights $w_{i\alpha\beta}$ that best reproduce the true redshifts 
$z_{\rm spec}$ in a spectroscopic validation set.

In our implementation, we use a network configuration $N_m:15:15:15:1$, 
which receives $N_m$ magnitudes and outputs a photo-z. 
We use hyperbolic tangent activation  functions in the hidden layers and a 
linear activation function for the output layer.

\end{document}